\documentclass[prd, preprint, 
  superscriptaddress,
  tightenlines,nofootinbib,
  eqsecnum,
  showpacs]{revtex4}

\usepackage{amsmath}
\usepackage{amsfonts}
\usepackage{amssymb}
\usepackage{bm}
\usepackage{hyperref}
\usepackage{mathrsfs}
\usepackage{graphicx}

\usepackage{ulem}
\usepackage[usenames]{color}

\allowdisplaybreaks

\DeclareSymbolFontAlphabet{\mathrsfs}{rsfs}
\DeclareMathAlphabet{\mathcal}{OMS}{cmsy}{m}{n}

\newcommand{\ud}{\mathrm{d}}

\newcommand{\beq}{\begin{equation}}
\newcommand{\eeq}{\end{equation}}

\newcounter{theorem} 

\begin{document}

\title{Equations of motion of self-gravitating $N$-body systems \\in the first post-Minkowskian approximation}

\author{Luc Blanchet}\email{luc.blanchet@iap.fr}
\affiliation{$\mathcal{G}\mathbb{R}\varepsilon{\mathbb{C}}\mathcal{O}$, Institut d'Astrophysique de Paris,\\ UMR 7095, CNRS, Sorbonne Universit{\'e}s \& UPMC Univ Paris 6,\\ 98\textsuperscript{bis} boulevard Arago, 75014 Paris, France}

\author{Athanassios S. Fokas}\email{tf227@cam.ac.uk}
\affiliation{Department of Applied Mathematics and Theoretical Physics,\\ University of Cambridge, Cambridge, CB3 0WA, U.K.}
\affiliation{Viterbi School of Engineering, University of Southern California,\\ Los Angeles, California, 90089-2560, U.S.A.}

\date{\today}

\begin{abstract} 
We revisit the problem of the equations of motion of a system of $N$ self-interacting massive particles (without spins) in the first post-Minkowskian (1PM) approximation of general relativity. We write the equations of motion, gravitational field and associated conserved integrals of the motion in a form suitable for comparison with recently published post-Newtonian (PN) results at the 4PN order. We show that the Lagrangian associated with the equations of motion in harmonic coordinates is a generalized one, and compute all the terms linear in $G$ up to 5PN order. We discuss the Hamiltonian in the frame of the center of mass and exhibit a canonical transformation connecting it to previous results directly obtained with the Hamiltonian formalism of general relativity. Finally we recover the known result for the gravitational scattering angle of two particles at the 1PM order.
\end{abstract}


\maketitle

\section{Introduction} 
\label{sec:intro}

The post-Minkowskian (PM) expansion is one of the most important approximation scheme in general relativity. Assuming that the gravitational field is weak, it expands in non-linearities around the Minkowski background, with small dimensionless expansion parameter $r_\text{S}/r$, where $r_\text{S}\sim G m/c^2$ is the Schwarzschild radius of the source, and no restriction on the slowness parameter $v/c$.\footnote{Here, $r$ denotes the typical size and $v$ the typical velocity of the source. Henceforth, small $n$PM terms $\sim (r_\text{S}/r)^n$ will be said to be of order $G^n$ in Newton'constant.} It has been developed starting from the 1916 Einstein paper (linearized gravity)~\cite{E16}, by many pionneering works from the 1960's on~\cite{Bertotti56,Havas57,BertottiP60,WG79,HG62,Port80,Westpf85,Dgef96}. Notably the approximation scheme has been systematically investigated~\cite{BertottiP60,ThK75,CTh77}, and the equations of motion of systems of particles have been obtained up to 2PM order (quadratic in $G$)~\cite{WG79,WH80,Westpf85,BeDD81} (in particular, the work~\cite{BeDD81} introduced a regularization procedure for treating ultra-violet divergences due to point particles at 2PM order).

On the other hand, the post-Newtonian (PN) approximation, or expansion in powers of $v/c$ (the other very important approximation scheme in general relativity), has played a crucial role in the data analysis of the recent gravitational wave events. While the gravitational waveform generated by compact (black-hole or neutron star) binary systems is known to 3.5PN order beyond the quadrupole formula~\cite{BlanchetLR}, the equations of motion have been derived to the 4PN order by means of two independent approaches: The canonical Hamiltonian formalism in Arnowitt-Deser-Misner (ADM) coordinates~\cite{JaraS12,JaraS13,DJS14,DJS15eob,JaraS15,DJS16}, and the Fokker Lagrangian in harmonic coordinates~\cite{BBBFMa,BBBFMb,BBBFMc,MBBF17,BBFM17}. In addition, the effective field theory approach has obtained partial results at 4PN order (with missing $G^4$ terms in Newton's constant)~\cite{FS4PN,FStail,GLPR16,Foffa5PN}.

The most well known physical application of the PM approximation is for unbound orbits, \textit{i.e.} the scattering problem with high relative velocities and small deviation angle. The gravitational scattering of two particles has been solved at the 2PM order~\cite{WG79,Port80,WH80,Westpf85}. More recently, Ledvinka, Sch{\"a}fer and Bi\v{c}ak~\cite{LSB08} published a closed form expression for the Hamiltonian of $N$ particles in the 1PM approximation. Very recently new works appeared on the gravitational scattering angle and the Hamiltonian of two point particles at the 2PM order and the link with the effective-one-body formalism~\cite{Dscatt16,Dscatt17}.

The aim of the present paper is to revisit the PM approximation in order to check some terms in the recent PN derivations of the equations of motion of compact binary systems in harmonic coordinates. For this purpose we restrict attention to the 1PM approximation, keeping only linear terms in $G$. Of course, the 1PM approximation is well known, but we find that some work is necessary in order to make possible the comparison with PN results. In particular, we need to explicitly perform the expansion of retardations consistently with the approximation, so as to recast the 1PM equations of motion into PN like form (valid up to any PN order but order $G$ only). In addition we want to provide closed form expressions for the associated conserved quantities like energy. 

Concerning the Lagrangian, the problem is not so straightforward because in harmonic coordinates the Lagrangian is a generalized one, depending not only on positions and velocities but also on the accelerations of the bodies. We shall not find a closed form expression for the harmonic coordinates Lagrangian at the 1PM order in the general case, but propose a method to obtain it at any PN order (and linear order in $G$). In the two body case with equal masses we give explicit expressions for the Lagrangian and Hamiltonian, and we shall connect, \textit{via} a canonical transformation, the Hamiltonian to the one obtained in~\cite{LSB08}. Finally we have been able to check all the terms linear in $G$ in many of the recently obtained 4PN results~\cite{BBBFMa,BBBFMb,BBBFMc,MBBF17,BBFM17}.

The plan of this paper is as follows. In Sec.~\ref{sec:eom} we present the derivation of the 1PM gravitational field and equations of motion of $N$ particles in PN like form. In Sec.~\ref{sec:consE} we obtain the conserved energy and linear momentum in closed analytic form. The special case of two equal masses is investivated in Sec.~\ref{sec:eqmass}. We investigate the (acceleration dependent) Lagrangian in harmonic coordinates in Sec.~\ref{sec:lag}, and present new formulas for computing it at any PN order. We discuss the corresponding Hamiltonian formalism in Sec.~\ref{sec:Ham}. Further material is relegated to some appendices: the scattering of two particles at 1PM order in App.~\ref{app:scatt} (we recover the known result); the rather academic case of bounded circular orbits and their stability in App.~\ref{app:circ}; and the terms of order $G$ in the harmonic coordinates Lagrangian up to 5PN order in App.~\ref{app:5PN}.

\section{Derivation of the equations of motion} 
\label{sec:eom}

In this paper we restrict attention to the \textit{linearized} Einstein field equations in harmonic coordinates. Thus, our theory is based on the linearized Einstein-Hilbert action of general relativity and on the standard linearized harmonic gauge-fixing term. We denote by $h^{\mu\nu}=\sqrt{-g}g^{\mu\nu}-\eta^{\mu\nu}$ the ``gothic'' metric deviation from Minkowski's metric, where $g^{\mu\nu}$ is the inverse of the usual covariant metric $g_{\rho\sigma}$ and $g=\text{det}(g_{\rho\sigma})$. Furthermore, we raise and lower indices using the Minkowski metric, for instance $h=\eta^{\mu\nu}h_{\mu\nu}$. Thus, we have
\begin{equation}\label{S}
S = \int \frac{\ud^4x}{c} \left[ \frac{c^4}{64\pi G} \,h^{\mu\nu}\Bigl(\eta_{\mu(\rho}\eta_{\sigma)\nu}-\frac{1}{2}\eta_{\mu\nu}\eta_{\rho\sigma}\Bigr) \Box h^{\rho\sigma} - \frac{1}{2} \Bigl( h_{\mu\nu}-\frac{1}{2}h \Bigr) T^{\mu\nu}\right]\,.
\end{equation}
Here $\Box$ is the flat space-time d'Alembertian operator, and $T^{\mu\nu}$ is the stress-energy tensor of matter fields in \textit{special relativity}, corresponding to a system of $N$ point particles without spins, namely 
\begin{equation}\label{Tmunu}
T^{\mu\nu}(x) = \sum_{a=1}^N m_a c^2 \int_{-\infty}^{+\infty} \ud\tau_a \,u_a^\mu u_a^\nu \delta^{(4)}(x-y_a)\,.
\end{equation}
The sum runs over the $N$ particles, $\delta^{(4)}$ is the four-dimensional Dirac distribution, $m_a$'s are the constant masses of particles, the world-lines of the particles are denoted by $y_a^\mu$, and the particles' special-relativistic proper time is $\tau_a$, so that $\ud\tau_a = \sqrt{- \eta_{\mu\nu} \ud y_a^\mu \ud y_a^\nu/c^2}$. The dimensionless four-velocities are defined by $c u_a^\mu=\ud y_a^\mu/\ud\tau_a$, and are normalized to $\eta_{\mu\nu}u_a^\mu u_a^\nu=-1$. The linearized harmonic gauge-fixed Einstein field equations deduced from the action~\eqref{S} are given by
\begin{equation}\label{fieldeq}
\Box h^{\mu\nu} = \frac{16\pi G}{c^4} T^{\mu\nu} \,.
\end{equation}

We solve the latter equations for the case of the explicit stress-energy tensor of particles given by~\eqref{Tmunu} by means of the standard Lienard-Wiechert integration procedure~\cite{Jackson}: adopting a parametrization of the trajectories by means of the coordinate time $t$, \textit{i.e.} such that $y_a^\mu(t)=\bigl(c t, \bm{y}_a(t)\bigr)$, the retarded time $t_a^\text{ret}$ on the trajectory $a$ which is associated with the propagation from the particle $a$ to the field point $x^\mu=(c t, \mathbf{x})$, is given by the implicit equation
\begin{equation}\label{retarded}
t_a^\text{ret}(\mathbf{x}, t) = t - \frac{1}{c}\bigl|\mathbf{x} - \bm{y}_a\bigl(t_a^\text{ret}(\mathbf{x}, t)\bigr)\bigr|\,,
\end{equation}
where $\vert\cdot\cdot\vert$ denotes the usual Euclidean norm. Denoting by $r_a^\text{ret} = \vert \mathbf{x} - \bm{y}_a(t_a^\text{ret})\vert$ the spatial interval, and by $\bm{n}_a^\text{ret} = \bigl(\mathbf{x} - \bm{y}_a(t_a^\text{ret})\bigr)/r_a^\text{ret}$ the unit spatial direction of propagation from the source point to the field point, the retarded solution of Eq.~\eqref{fieldeq} is given by 
\begin{equation}\label{field0}
h^{\mu\nu}(\bm{x}, t) = - \frac{4}{c^2} \sum_a \frac{G m_a \,u_a^\mu u_a^\nu}{r_a^\text{ret} \,\gamma_a \bigl(1 - \bm{n}_a^\text{ret}\cdot\bm{v}_a/c\bigr)}\,.
\end{equation}
The factor
\begin{equation}\label{redshift}
k^a_\mu u_a^\mu =\gamma_a\Bigl(-1+\frac{\bm{n}_a^\text{ret}\cdot\bm{v}_a}{c}\Bigr)\,,
\end{equation}
which occurs in the denominator of~\eqref{field0}, is the standard redshift factor, where $k_a^\mu$ is the Minkowski null vector between source and field points, given by $k_a^\mu = [x^\mu-y_a^\mu(t_a^\text{ret})]/r_a^\text{ret} = (1,\bm{n}_a^\text{ret})$, where $\gamma_a=u_a^0$ is the usual Lorentz factor, and the ordinary coordinate velocity is defined by $v_a^\mu=c u_a^\mu/\gamma_a=(c, \bm{v}_a)$, with $\bm{n}_a^\text{ret}\cdot\bm{v}_a$ denoting the ordinary Euclidean scalar product. 

The velocities in Eq.~\eqref{field0} should also be computed at the retarded time $t_a^\text{ret}$. However, here we shall repeatedly use the fact that the accelerations are of order $G$ and therefore their contributions in Eq.~\eqref{field0} will be of order $G^2$, and hence can be neglected within the present approximation which is strictly confined to the linear order in $G$. Thus, in all our computations we can always assume that the four velocity $u_a^\mu$ and Lorentz factor $\gamma_a$ are constant. Furthermore, neglecting terms of order $G$, we can solve the retardation equation~\eqref{retarded}, to reexpress the retarded time $t_a^\text{ret}$, distance $r_a^\text{ret}$, and direction $\bm{n}_a^\text{ret}$, in terms of their values at the current time $t$, \textit{i.e.} the ``\textit{instantaneous}'' distance $r_a=\vert\mathbf{x}-\bm{y}_a(t)\vert$ and direction $\bm{n}_a = [\mathbf{x}-\bm{y}_a(t)]/r_a$. In this way, we find
\begin{subequations}\label{solveretard}
\begin{align}
r_a^\text{ret} &= \gamma_a\,r_a\left[\sqrt{1 + (n_a u_a)^2} + (n_a u_a)\right]\,,\\\bm{n}_a^\text{ret} &= \frac{\bm{v}_a}{c} + \frac{\bm{n}_a}{\gamma_a}\left[\sqrt{1 + (n_a u_a)^2} - (n_a u_a)\right]\,.
\end{align}
\end{subequations}

In what follows, for convenience of notation we use parenthesis to denote the ordinary scalar product, \textit{i.e.} $(n_a u_a) = \bm{n}_a\cdot\bm{u}_a = \gamma_a (n_a v_a)/c$. Substituting these formulas into Eq.~\eqref{field0} we find that the equivalent instantaneous expression of the field takes the form
\begin{equation}\label{field}
h^{\mu\nu} = - \frac{4}{c^2} \sum_a \frac{G m_a \,u_a^\mu u_a^\nu}{r_a \sqrt{1 + (n_a u_a)^2}}\,.
\end{equation}
%

Next, in preparation for the derivation of the geodesic equation, we compute the space-time derivatives of the field, which can be easily achieved using the explicit expression~\eqref{field}. Recalling that within the present approximation the four-velocities are constant, we find
\begin{equation}\label{derfield}
\partial_\rho h^{\mu\nu} = \frac{4}{c^2} \sum_a \frac{G m_a \,u_a^\mu u_a^\nu}{r_a^2 \left[1 + (n_a u_a)^2\right]^{3/2}}\Bigl( n^a_\rho + u^a_\rho (n_a u_a)\Bigr)\,.
\end{equation}
Here we have extended the definition of the unit direction $\bm{n}_a$ by posing $n_a^\mu = (0, \bm{n}_a)$; also, we use $u_a^\mu = \gamma_a (1, \bm{v}_a/c)$. An immediate check of the result~\eqref{derfield} is that the divergence of $h^{\mu\nu}$ is seen to be zero: $H^\mu=\partial_\nu h^{\mu\nu}=0$. This is consistent with the harmonic-coordinates field equations~\eqref{fieldeq} together with the conservation law of the matter tensor which is identical with the conservation law in special relativity, namely $\partial_\nu T^{\mu\nu}=0$.

The expressions~\eqref{field} and~\eqref{derfield} are valid everywhere except at the singular location of the particles. Nevertheless, we can easily extend the validity of these equations by using a self-field regularization, which deals with the infinite self-field of particles. For the purpose, it is sufficient to simply discard the self-field contribution from the sum of the particles. Therefore, at the particle $a$, the expression~\eqref{derfield} becomes
\begin{equation}\label{derfielda}
(\partial_\rho h^{\mu\nu})_a = \frac{4}{c^2} \sum_{b\not= a} \frac{G m_b \,u_b^\mu u_b^\nu}{r_{ab}^2 \bigl[1 + (n_{ab} u_b)^2\bigr]^{3/2}}\Bigl( n^{ab}_\rho + u^{ab}_\rho (n_{ab} u_b)\Bigr)\,,
\end{equation}
where now the sum runs over all particles except $a$, and where $r_{ab}=\vert\bm{y}_a-\bm{y}_b\vert$ and $n_{ab}^0=0$, $n_{ab}^i=[\bm{y}_a - \bm{y}_b]/r_{ab}$. The equation of motion is obtained by just inserting the latter expression into the geodesic equation. At linearized order, which is consistent with our approximation, it is given by
\begin{equation}\label{geodesic}
\frac{\ud u_a^\mu}{\ud\tau_a} = \frac{c}{2} u_a^\rho u_a^\sigma \Bigl[ (\partial^\mu k_{\rho\sigma})_a - 2 (\partial_\rho
  k^\mu_\sigma)_a - 
  u_a^\mu u_a^\lambda (\partial_\lambda k_{\rho\sigma})_a \Bigr]
\,,
\end{equation}
where we have used the trace-reversed variable $k^{\mu\nu} = \frac{1}{2}\eta^{\mu\nu}h - h^{\mu\nu}$, that represents the linear perturbation in the ordinary covariant metric $g_{\mu\nu}$. Hence, we find that the complete equation of motion at the 1PM approximation is given by
\begin{align}\label{eom1}
\frac{\ud u_a^\mu}{\ud\tau_a} = - \frac{1}{c} \sum_{b\not= a} \frac{G m_b}{r_{ab}^2 \bigl[1 + (n_{ab} u_b)^2\bigr]^{3/2}}\biggl[ & (2\epsilon_{ab}^2-1)n_{ab}^\mu + (2\epsilon_{ab}^2+1)\Bigl( - (n_{ab} u_a) + \epsilon_{ab} (n_{ab} u_b)\Bigr) u_a^\mu \nonumber\\ &+ \Bigl( 4 \epsilon_{ab} (n_{ab} u_a) - (2\epsilon_{ab}^2+1) (n_{ab} u_b) \Bigr) u_b^\mu\biggr] \,.
\end{align}
We have used $\epsilon_{ab} = - (u_a u_b)$ as a shorthand notation;\footnote{This notation agrees with the energy function used for two-body problems in particle physics, in particular $$\epsilon_{ab} = \frac{s_{ab}-m_a^2-m_b^2}{2m_a m_b} = - \frac{(p_a p_b)}{m_a m_b c^2}\,,$$
where $s_{ab} = - (p_a+p_b)^2/c^2$ is the Mandelstam variable, see \textit{e.g.}~\cite{Dgef96,Dscatt16,Dscatt17}.} parenthesis represent four-dimensional scalar products, for instance $(n_{ab} u_a) = n^{ab}_\mu u_a^\mu = \gamma_a \bm{n}_{ab}\cdot\bm{v}_a/c$
and $\epsilon_{ab} = \gamma_a \gamma_b (1-\bm{v}_a\cdot\bm{v}_b/c^2)$. Using~\eqref{eom1} we can verify that $(u_a \frac{\ud u_a}{\ud\tau_a})=0$ which is of course, the consequence of the normalization $(u_a u_a) = -1$.

Finally, we present the non-covariant form of the equations of motion, introducing the ordinary velocity and acceleration fields and the relevant Lorentz factors. It turns out that this form is relatively simple, 
\begin{equation}\label{eom2}
\frac{\ud \bm{v}_a}{\ud t} = - \gamma_a^{-2}\sum_{b\not= a} \frac{G m_b}{r_{ab}^2 \, y_{ab}^{3/2}}\biggl[ (2\epsilon_{ab}^2-1)\bm{n}_{ab} + \gamma_b \Bigl( - 4 \epsilon_{ab} \gamma_a (n_{ab} v_a) + (2\epsilon_{ab}^2+1) \gamma_b (n_{ab} v_b) \Bigr) \frac{\bm{v}_{ab}}{c^2}\biggr]\,,
\end{equation}
where for convenience of notation we use $y_{ab} = 1 + \gamma_b^2(n_{ab} v_b)^2/c^2$ (notice that $y_{ab}\not= y_{ba}$). We note that the second term is proportional to the relative velocity $\bm{v}_{ab}=\bm{v}_{a}-\bm{v}_{b}$, and that there are no terms proportional to individual velocities $\bm{v}_{a}$ or $\bm{v}_{b}$.

We have verified that the formula~\eqref{eom2} exactly reproduces the post-Newtonian results in the case of two particles ($N=2$) at linear order in $G$ and up to the 4PN order~\cite{BBBFMc,BBFM17}.

\section{Conserved energy and linear momentum}
\label{sec:consE}

The equations of motion at the 1PM approximation, \textit{i.e.} Eqs.~\eqref{eom1} or~\eqref{eom2} are \textit{conservative}, and thus admit conserved integrals of energy, angular momentum and linear momentum. Indeed, the gravitational radiation reaction dissipative effects in the equations of motion are at least of second order in $G$, see, \textit{e.g.}~\cite{BlanchetLR}. To find the conserved energy we proceed in a standard way, namely, we form the combination $W=\sum_a m_a\gamma_a^3 (v_a a_a)$, where $\bm{a}_a$ denotes the acceleration~\eqref{eom2}. By definition, $W$ is the time-derivative of the special-relativistic total energy, $W=\frac{\ud}{\ud t}[\sum_a m_a \gamma_a c^2]$. Also, after replacing $\bm{a}_a$ \textit{via} Eq.~\eqref{eom2}, we need to rewrite the resulting expression in the form of the total time-derivative of some potential up to order $G$, say $W= -\frac{\ud V}{\ud t}$. Then, the conserved energy $E$, satisfying $\frac{\ud E}{\ud t}=0$, is given by 
\begin{equation}\label{Edef}
E = \sum_a m_a \gamma_a c^2 + V\,.
\end{equation}

We find this energy in two steps. First, a part of the terms in $W$ can be readily integrated thanks to the easily checked identity
\begin{equation}\label{formula}
\frac{\ud}{\ud t}\left(\frac{1}{r_{ab} \, y_{ab}^{1/2}}\right) = \frac{\frac{\gamma_b}{\gamma_a}\epsilon_{ab} (n_{ab}v_b) - (n_{ab}v_a)}{r_{ab}^2 \, y_{ab}^{3/2}} \,,
\end{equation}
where $y_{ab}$ was defined \textit{via} the equation $y_{ab} = 1 + \gamma_b^2(n_{ab} v_b)^2/c^2$ and we recall that $\epsilon_{ab}=\gamma_a \gamma_b[1-(v_av_b)/c^2]$. In this way, several terms occuring in the potential $V$ are obtained,
\begin{align}\label{V0}
V &= \sum_a \sum_{b\not= a}G m_a m_b \biggl[ \gamma_a\frac{2\epsilon_{ab}^2+1-4\frac{\gamma_b}{\gamma_a}\epsilon_{ab}}{r_{ab}\,y_{ab}^{1/2}} + \frac{\gamma_b^2}{\gamma_a}\left(2\epsilon_{ab}^2-1\right) (D_{ab}v_b)\biggr]\,.
\end{align}
However, we find that the last term corresponds to a more complicated structure of the energy, and is given by the scalar product $(D_{ab}v_b)=\bm{D}_{ab}\cdot\bm{v}_b=D_{ab}^i v_b^i$, between the velocity $v_b^i$ (which is constant) and some elementary solution of the equation
\begin{equation}\label{dDab}
\frac{\ud D^i_{ab}}{\ud t} = \frac{n^i_{ab}}{r_{ab}^2\,y_{ab}^{3/2}}\,.
\end{equation}

We have obtained an analytic closed-form solution of the equation~\eqref{dDab} using the following approach: recalling that within our approximation the velocities are constant and trajectories are straight lines, we can always obtain the unit direction vector $n^i_{ab}=(y_a^i-y_b^i)/r_{ab}$ as a simple function of the relative distance $r_{ab}$, namely
\begin{equation}\label{nab}
n^i_{ab} = \frac{v^i_{ab}}{v_{ab}^2}\sqrt{v_{ab}^2 - \frac{C_{ab}^2}{r_{ab}^2}}+\frac{\mu^i_{ab}}{r_{ab}}\,,
\end{equation}
where $C_{ab}$ and the three-vector $\bm{\mu}_{ab}=(\mu^i_{ab})$ are \textit{constants of the motion}. The first constant, $C_{ab}$, is nothing but the specific angular momentum (\textit{i.e.}, the angular momentum per unit mass) of the relative motion of the two particles $a$ and $b$; the second constant, $\bm{\mu}_{ab}$, is related to the initial positions at time $t=0$ of the two particles on their straight-line trajectories.\footnote{More precisely, let the trajectories be $\bm{y}_a = \bm{v}_a t + \bm{y}_a^0$, and pose $\bm{y}_{ab}^0 = \bm{y}_a^0 - \bm{y}_b^0$ for the initial relative position at $t=0$. Then we have
$$\bm{\mu}_{ab} = \bm{y}_{ab}^0 - \frac{(y_{ab}^0v_{ab})\bm{v}_{ab}}{v_{ab}^2}\,.$$
Note that this vector is perpendicular to the relative velocity: $(\mu_{ab}v_{ab})=0$.} We note that $C_{ab}=C_{ba}$ whereas $\bm{\mu}_{ab}=-\bm{\mu}_{ba}$. From Eq.~\eqref{nab} we obtain the scalar products $(n_{ab} v_a)$ and $(n_{ab} v_b)$ as
\begin{equation}\label{nabva}
(n_{ab} v_a) = \frac{(v_{ab}v_a)}{v_{ab}^2}\sqrt{v_{ab}^2 - \frac{C_{ab}^2}{r_{ab}^2}}+\frac{(\mu_{ab}v_{a})}{r_{ab}}\,,
\end{equation}
together with $a\leftrightarrow b$. We recall our notations $(v_{ab}v_a)=v_a^2-(v_a v_b)$, $(v_{ab}v_b)=-v_b^2+(v_a v_b)$ and $v_{ab}^2=v_a^2-2(v_a v_b)+v_b^2$. Next we obtain, using Eq.~\eqref{nabva}, the time derivative of the relative distance $r_{ab}$ as
\begin{equation}\label{drab}
\frac{\ud r_{ab}}{\ud t} = (n_{ab} v_a) - (n_{ab} v_b) = \sqrt{v_{ab}^2 - \frac{C_{ab}^2}{r_{ab}^2}}\,.
\end{equation}
The above formulae are valid in that portion of the trajectories for which $r_{ab}$ increases. Equation~\eqref{drab} gives us a one-to-one relationship between time $t$ and the distance $r_{ab}$, and allows one to perform a change of variable in Eq.~\eqref{dDab} from the variable $t$ to the distance $r_{ab}$ itself, and hence we obtain the ordinary differential equation
\begin{equation}\label{ODE}
\frac{\ud D^i_{ab}}{\ud r_{ab}} = \frac{\frac{v^i_{ab}}{v_{ab}^2} + \frac{\mu^i_{ab}}{\sqrt{v_{ab}^2 r_{ab}^2 - C_{ab}^2}}}{r_{ab}^2\left(1+\frac{\gamma_b^2}{c^2}\left[\frac{(v_{ab}v_b)}{v_{ab}^2}\sqrt{v_{ab}^2 - \frac{C_{ab}^2}{r_{ab}^2}}+\frac{(\mu_{ab}v_b)}{r_{ab}}\right]^2\right)^{3/2}}\,.
\end{equation}
This equation can be integrated in closed analytic form:
\begin{equation}\label{closedform}
D^i_{ab} = \frac{1}{C_{ab}^2+\frac{\gamma_b^2}{c^2}\left[(\mu_{ab}v_b)^2v_{ab}^2+C_{ab}^2\frac{(v_{ab}v_b)^2}{v_{ab}^2}\right]}\times\frac{\mu^i_{ab}\sqrt{v_{ab}^2r_{ab}^2 - C_{ab}^2} - C_{ab}^2\frac{v^i_{ab}}{v_{ab}^2}}{r_{ab} \left(1+\frac{\gamma_b^2}{c^2}\left[\frac{(v_{ab}v_b)}{v_{ab}^2}\sqrt{v_{ab}^2 - \frac{C_{ab}^2}{r_{ab}^2}}+\frac{(v_{ab}v_b)}{r_{ab}}\right]^2\right)^{1/2}} \,.
\end{equation}
We note that the first factor in the above equation, within our approximation, is a pure constant. Finally, this form as it stands is not very useful, and we express the constants $C_{ab}$ and $\mu^i_{ab}$ in terms of their earlied expressions as functions of the original variables $n^i_{ab}$, $v^i_{ab}$, the scalar products $(n_{ab}v_a)$ and $(n_{ab}v_b)$ and $(n_{ab}v_{ab})=\dot{r}_{ab}\equiv\ud r_{ab}/\ud t$. This gives the following interesting structure for the most ``complicated'' term in the energy:
\begin{equation}\label{Dabsol}
D^i_{ab} = \frac{\bigl[\dot{r}_{ab}+\frac{\gamma_b^2}{c^2}(n_{ab}v_{b})(v_{ab}v_{b})\bigr]n^i_{ab}-y_{ab}\,v^i_{ab}}{r_{ab}\,y_{ab}^{1/2}\left[\bigl(v_{ab}^2-\dot{r}_{ab}^2\bigr)y_{ab}+\frac{\gamma_b^2}{c^2}\bigl(\dot{r}_{ab}(n_{ab}v_{b})-(v_{ab}v_b)\bigr)^2\right]}\,,
\end{equation}
plus an integration constant which can be absorbed into the definition of $E$. Hence, combining Eqs.~\eqref{V0} and~\eqref{Dabsol}, we have a closed-form expression for the conserved energy~\eqref{Edef}, thoroughly given by the expression for the potential $V$ at order $G$:
\begin{align}\label{V}
V &= \sum_a \sum_{b\not= a}\frac{G m_a m_b}{r_{ab}\,y_{ab}^{1/2}} \Biggl\{ \gamma_a\Bigl(2\epsilon_{ab}^2+1-4\frac{\gamma_b}{\gamma_a}\epsilon_{ab}\Bigr) \\& \qquad\qquad + \frac{\gamma_b^2}{\gamma_a}\bigl(2\epsilon_{ab}^2-1\bigr) \frac{\dot{r}_{ab}(n_{ab}v_b)-(v_{ab}v_b)}{\bigl(v_{ab}^2-\dot{r}_{ab}^2\bigr)y_{ab}+\frac{\gamma_b^2}{c^2}\bigl(\dot{r}_{ab}(n_{ab}v_{b})-(v_{ab}v_b)\bigr)^2}\Biggr\}\nonumber\,.
\end{align}
We have verified that this energy reproduces in the post-Newtonian limit for two particles the 4PN results up to order $G$~\cite{BBBFMc,BBFM17}. 

We apply next the same method to the integral of linear momentum. Namely, we start by forming the combination $\frac{\ud}{\ud t}[\sum_a m_a \gamma_a \bm{v}_a] \equiv \sum_a m_a \gamma_a[ \bm{a}_a+ \gamma_a^2 (v_a a_a)\bm{v}_a/c^2]$, since $\sum_a m_a \gamma_a \bm{v}_a$ represents the total linear momentum in special relativity. Then, we replace the accelerations by the equations of motion, and are able to transform the result into a total time-derivative thanks to the integration formulas~\eqref{formula} and, most importantly,~\eqref{Dabsol}. As a result we obtain the conserved total linear momentum, such that $\ud \bm{P}/\ud t = \bm{0}$, as
\begin{equation}\label{Pidef}
P^i = \sum_a m_a \gamma_a v_a^i + \Pi^i\,,
\end{equation}
in which the terms of order $G$ are given by
\begin{align}\label{Pi}
\Pi^i &= \sum_a \sum_{b\not= a}\frac{G m_a m_b}{r_{ab}\,y_{ab}^{1/2}} \Biggl\{ \bigl(2\epsilon_{ab}^2+1\bigr) \gamma_a \frac{v_a^i}{c^2} - 4 \epsilon_{ab} \gamma_b \frac{v_b^i}{c^2} \\& \qquad\qquad\qquad + \frac{2\epsilon_{ab}^2-1}{\gamma_a} \Bigl(\delta^{ij}+\frac{\gamma_b^2}{c^2}v_b^i v_b^j\Bigr) \frac{\bigl[\dot{r}_{ab}+\frac{\gamma_b^2}{c^2}(n_{ab}v_{b})(v_{ab}v_{b})\bigr]n^j_{ab}-y_{ab}v^j_{ab}}{\bigl(v_{ab}^2-\dot{r}_{ab}^2\bigr)y_{ab}+\frac{\gamma_b^2}{c^2}\bigl(\dot{r}_{ab}(n_{ab}v_{b})-(v_{ab}v_b)\bigr)^2}\Biggr\}\nonumber\,.
\end{align}

As is well known, the main physical situation meaningfully described by the PM approximation is the scattering of particles. The scattering of two particles has been worked out long ago up to the 2PM order~\cite{WG79,Port80,Westpf85}, see Eq.~(4.78) in~\cite{Westpf85}. In the Appendix~\ref{app:scatt} we shall check that the previous formalism, in particular the integration formulas~\eqref{closedform}--\eqref{Dabsol}, correctly recovers the known result for the scattering of two particles at the 1PM order. 

\section{The two-body case with equal masses}
\label{sec:eqmass}

In this section we discuss the equations of motion in the case of two bodies ($N=2$) with equal masses. For equal masses we necessarily have $\bm{v}_1^i = - \bm{v}_2^i$ in the frame of the center of mass, which is an important simplification of the general case. We denote
\begin{equation}\label{eqmassnot}
m_1 = m_2 = \frac{m}{2} \,, \quad \bm{v}_1^i = - \bm{v}_2^i = \frac{\bm{v}}{2} \,, \quad\text{and}\quad \gamma_1=\gamma_2=\gamma\,,
\end{equation}
where $m=m_1+m_2$ is the total mass, $\bm{v}=\bm{v}_1-\bm{v}_2$ is the relative velocity, and $\gamma=(1-\frac{v^2}{4 c^2})^{-1/2}$ is the common Lorentz factor of the two bodies. Notice that $\epsilon_{12}=2\gamma^2-1$. We also pose $r=r_{12}=\vert\bm{y}_1-\bm{y}_2\vert$ and $\bm{n}=\bm{n}_{12}=(\bm{y}_1-\bm{y}_2)/r$, such that $\dot{r}=(nv)$. We also denote, consistently with our previous notation in Sec.~\ref{sec:eom}, $y=1+\frac{\gamma^2(nv)^2}{4 c^2}$ (note that $y=y_{12}=y_{21}$ for equal masses). 

Inserting~\eqref{eqmassnot} in the basic equations of motion~\eqref{eom1}--\eqref{eom2}, it follows that the motion of two equal masses in the 1PM approximation is characterized by the single equation\footnote{Or, equivalently,$$\frac{\ud v^i}{\ud t} = - \frac{G m}{r^2 y^{3/2}}\left[ \Bigl(8 \gamma^2-8+\frac{1}{\gamma^2}\Bigr)\,n^i - \Bigl(4\gamma^4-\frac{1}{2}\Bigr) (nv)\,\frac{v^i}{c^2} \right] \,.$$}
\begin{align}\label{singleeom}
\frac{\ud}{\ud\tau}(\gamma \bm{v}) = -\frac{Gm}{r^{2}y^{3/2}}\left[
\bigl(8\gamma^{4}-8\gamma^{2}+1\bigr)\bm{n}-\frac{\gamma^{2}}{4}\bigl(16\gamma^{6}-8\gamma^{4}+6\gamma^{2}-1\bigr)\frac{\dot{r} \bm{v}}
{c^{2}}\right]\,.
\end{align}
Recalling the definition of the force, we obtain
\begin{align}\label{force}
\bm{f} = m\frac{\ud}{\ud t}(\gamma\bm{v}) = -\frac{Gm^{2}}{
r^{2}y^{3/2}}
\left[\bigl(8\gamma^{3}-8\gamma+\gamma^{-1}\bigr)\bm{n}-
\frac{1}{4}\bigl(16\gamma^{7}-8\gamma^{5}+6\gamma^{3}-\gamma\bigr)\frac{\dot{r} \bm{v}}
{c^{2}}\right]\,.
\end{align}
%
Of course, in the small velocity limit, we recover the form of the usual Newtonian gravitational force $\bm{f}_\text{N} = -\frac{G m^{2}}{
r^{2}} \bm{n}$.

Next we derive the evolution equation for $\gamma^{2}$. We use the identity $\frac{\ud}{\ud t}(\gamma \bm{v}\cdot\gamma \bm{v})
=2\gamma \bm{v}\cdot \frac{\ud}{\ud t} (\gamma \bm{v})$ together with $ \gamma \bm{v}\cdot\gamma \bm{v}=
\gamma^{2}v^{2}=4c^{2}(\gamma^{2}-1)$, to arrive at
\begin{equation}\label{dgammadt}
\frac{\ud\gamma^{2}}{\ud t}=\frac{8 Gm \gamma^{2}}{r^{2}y^{3/2}c^{2}}\dot{r}
\left(\gamma^{6}-\frac{3}{2}\gamma^{4}+\frac{3}{8}\gamma^{2}
+\frac{1}{16}\right)\,,
\end{equation}
or equivalently, changing the variable $t\longrightarrow r$,
\begin{equation}\label{dgammadr}
\frac{\ud\gamma^{2}}{\ud r}=\frac{8G m}{r^{2}y^{3/2}c^{2}}
\biggl(\gamma^{8}-\frac{3}{2}\gamma^{6}+\frac{3}{8}\gamma^{4}
+\frac{\gamma^{2}}{16}\biggr)\,.
\end{equation}
Similarly we derive also the evolution equation for $y$. We consider the following identity $\frac{\ud}{\ud t}(\gamma \dot{r}) = \frac{\ud \bm{n}}{\ud t}\cdot\gamma\bm{v}+\bm{n}\cdot\frac{\ud}{\ud t}(\gamma \bm{v})$ and the fact that $v^2=\dot{r}^2+r \bm{v}\cdot\frac{\ud \bm{n}}{\ud t}$, which gives, after insertion of the equation of motion~\eqref{singleeom},
\begin{align}\label{dydt}
\frac{\ud y}{\ud t}=\dot{r}\left\{\frac{2}{r}(\gamma^{2}-y)+\frac{8 G m}{r^{2}c^{2}
y^{1/2}}
\biggl(\gamma^{6}-\frac{\gamma^{4}}{2}+\frac{3\gamma^{2}}{8}-\frac{1}{16}\biggr)
-\frac{8Gm}{r^{2}c^{2}y^{3/2}}\biggl(\gamma^{6}
-\frac{\gamma^{2}}{8}\biggr)\right\}\,.
\end{align}
Hence, again with changing variable from $t$ to $r$, we obtain 
\begin{align}\label{dydr}
\frac{\ud y}{\ud r}+\frac{2}{r}y-\frac{2}{r}\gamma^{2}=\frac{8Gm}{r^{2}c^{2}y^{3/2}}
\left[y\biggl(\gamma^{6}-\frac{\gamma^{4}}{2}+\frac{3}{8}\gamma^{2}-\frac{1}{16}\biggr)
-\biggl(\gamma^{6}-\frac{\gamma^{2}}{8}\biggr)\right].
\end{align}
Equations~\eqref{dgammadr} and~\eqref{dydr} form a system of two coupled ordinary differential equations which determine the two unknown functions $\gamma$ and $y$ in terms of $r$. Actually, it will be shown below that it is possible to express explicitly $y$ in term of $\gamma$ and $r$. Before deriving this expression we will derive the expressions for the conserved energy and conserved angular momentum.

The energy $E$ is defined by~\eqref{Edef} hence $E=mc^{2}\gamma+V$ for the two equal masses case. Hence, conservation of energy implies $\frac{\ud V}{\ud r}=-mc^{2}\frac{\ud \gamma}{\ud r}$. Thus, using equation~\eqref{dgammadr} we find
\begin{equation}\label{dVdr}
\frac{\ud V}{\ud r} = -\frac{8Gm^{2}}{r^{2}y^{3/2}}\biggl(
\gamma^{7}-\frac{3}{2}\gamma^{5}+\frac{3}{8}\gamma^{3}+\frac{\gamma}{16}\biggr)\,.
\end{equation}
In order to integrate the above equation  we first observe that since $\ud\gamma/\ud t=\mathcal{O}(G)$, $\gamma$ can be treated as constant. Furthermore, we will employ the identity
\begin{equation}\label{identity}
\frac{\ud}{\ud r}\left(\frac{1}{r y^{1/2}}\right) = -\frac{\gamma^{2}}{r^{2}y^{3/2}}+\mathcal{O}(G)\,,
\end{equation}
which is a consequence of Eq.~\eqref{dydr} when neglecting terms of order $G$. The integration of Eq.~\eqref{dVdr} is then straightforward and we obtain
\begin{equation}\label{intV}
V=\frac{G m^{2}}{r y^{1/2}}\bigl(2\gamma^2-1\bigr)\Bigl(2\gamma^{3}-2\gamma-\frac{1}{4\gamma}\biggr)\,.
\end{equation}
This result is perfectly consistent with the two equal-mass case of the general analysis done in the previous section, see Eq.~\eqref{V}.

The angular momentum $\bm{J}$ is given by the usual special-relativistic expression plus a 1PM correction $\bm{K}$ of order $G$, \textit{i.e.} $\bm{J}=m_{1}\gamma_{1}\,\bm{y}_{1}\times \bm{v}_{1}+m_{2}\gamma_{2}\,\bm{y}_{2}\times \bm{v}_{2}+\bm{K}$, which gives for the case at hands (denoting $\bm{C}=\bm{x}\times\bm{v}$)
\begin{equation}\label{Jdef}
\bm{J} = \frac{m \gamma}{4}\,\bm{C} +\bm{K}\,. 
\end{equation}
Thus $\frac{\ud \bm{K}}{\ud t} = -\frac{m}{4}\,\frac{\ud}{\ud t}(\gamma\bm{C}) = -\frac{1}{4}\,\bm{x}\times\bm{f}$, and then replacing the force by the right-hand side (RHS) of Eq.~\eqref{force} we obtain
\begin{equation*}\label{dKdt}
\frac{\ud \bm{K}}{\ud t}=-\bm{C}\,\frac{Gm^{2}}{c^{2}r^{2}y^{3/2}}\biggl(2\gamma^{7}-\gamma^{5}+2\gamma^{3}-\frac{\gamma}{8}\biggr)\,.
\end{equation*}
We can integrate this equation treating both $\gamma$ and $\bm{C}$ as constants to order $G$, and noticing that $C=2c r\sqrt{1-y/\gamma^2}$ in our notation. Thus, using also the identity~\eqref{identity}, we find
\begin{equation}\label{Kint}
\bm{K} = \bm{C}\,\frac{Gm^{2}}{c^{2}ry^{1/2}}\biggl(\gamma^{5}-\frac{\gamma^{3}}{2}+\frac{3}{8}\gamma-\frac{1}{16\gamma}\biggr)\,.
\end{equation}

Let $x=\gamma^{2}$. It is remarkable that the polynomial in $x$ appearing in the RHS of~\eqref{dgammadr} can be written in terms of the two polynomials appearing in the RHS of~\eqref{dydr}:
\begin{equation}\label{polynomial}
x^{4}-\frac{3}{2}x^{3}+\frac{3}{8}x^{2}+\frac{x}{16}=
x\biggl(x^{3}-\frac{x^{2}}{2}+\frac{3}{8}x-\frac{1}{16}\biggr)-\biggl(x^{3}-\frac{x}{8}\biggr)\,.
\end{equation}
Thus, Eq.~\eqref{dgammadr} can be rewritten in the form
\begin{equation}\label{dxdr}
\frac{\ud x}{\ud r} = \frac{8G m}{r^{2}y^{3/2}c^{2}}\left[x
\biggl(x^{3}-\frac{x^{2}}{2}+\frac{3}{8}x-\frac{1}{16}\biggr) - \biggl(x^{3}-\frac{x}{8}\biggr)\right]\,.
\end{equation}
Subtracting Eqs.~\eqref{dydr} and~\eqref{dxdr} we find
\begin{equation}\label{dxydr}
\frac{\ud}{\ud r}\bigl(x-y\bigr) + \frac{2}{r}\bigl(x-y\bigr) = \frac{8 G m}{r^{2}y^{3/2}c^{2}}
\bigl(x-y\bigr)\biggl(x^{3}-\frac{x^{2}}{2}+\frac{3}{8}x-\frac{1}{16}\biggr)\,,
\end{equation}
hence
\begin{equation}\label{dxydr2}
\frac{\frac{\ud}{\ud r}[r^{2}(x-y)]}{r^{2}(x-y)} = \frac{8 G m}{r^{2}y^{3/2}c^{2}} \biggl(x^{3}-\frac{x^{2}}{2}+\frac{3}{8}x-\frac{1}{16}\biggr)\,.
\end{equation}
Dividing the above equation and~\eqref{dgammadr} we find
\begin{equation}\label{dxydr3}
\frac{\ud[r^{2}(x-y)]}{r^{2}(x-y)}=\frac{x^{3}-\frac{x^{2}}{2}+\frac{3}{8}x-\frac{1}{16}}
{x^{4}-\frac{3x^{3}}{2}+\frac{3}{8}x^{2}+\frac{x}{16}}\,\ud x\,.
\end{equation}
Using the fact that the denominator of the RHS of~\eqref{dxydr3} factorizes,\footnote{Namely,
$$x^{4}-\frac{3x^{3}}{2}+\frac{3}{8}x^{2}+\frac{x}{16}=x\left(x-\frac{1}{2}\right)
\left(x^{2}-x-\frac{1}{8}\right)\,.$$} we can finally integrate~\eqref{dxydr3} in the nice form
\begin{equation}\label{nice}
r^{2}\bigl(x-y\bigr) = k \,\frac{\left(x^{2}-x-\frac{1}{8}\right)^{4/3}}
{x\left(x-\frac{1}{2}\right)^{2/3}}\,,
\end{equation}
where $k$ is a constant. In summary, the basic equation for $\gamma$ is Eq.~\eqref{dgammadr}, where $y$ is given by
\begin{equation}\label{nice2}
y = \gamma^{2}- k\,\frac{\left(\gamma^{4}-\gamma^{2}-\frac{1}{8}\right)^{4/3}}{r^{2}\gamma^{2}\left(\gamma^{2}-\frac{1}{2}\right)^{2/3}}\,.
\end{equation}

Of course the equation~\eqref{nice2} is physically meaningful only at the 1PM order, \textit{i.e.} at leading order in $G$, and the constant $k$ is not independent from the two fundamental integrals of motion $E$ and $J$. An easy calculation shows that
\begin{equation}\label{k}
k = \frac{4J^2c^2}{E^2} \,\frac{\bar{\gamma}^4\left(\bar{\gamma}^2-\frac{1}{2}\right)^{2/3}}{\left(\bar{\gamma}^{4}-\bar{\gamma}^{2}-\frac{1}{8}\right)^{4/3}}\,,
\end{equation}
where $\bar{\gamma}=\frac{E}{m c^2}$ denotes the constant Lorentz factor of special relativity. The integral of motion~\eqref{nice2} expanded at order $G$ is equivalent to saying that $\gamma=\bar{\gamma}+\delta\gamma$ where
\begin{equation}\label{deltagamma}
\delta\gamma = - \frac{G m}{r \bar{y}^{1/2}}\bigl(2\bar{\gamma}^2-1\bigr)\Bigl(2\bar{\gamma}^{3}-2\bar{\gamma}-\frac{1}{4\bar{\gamma}}\Bigr)\,,
\end{equation}
in agreement with the earlier result~\eqref{intV}. Note that here, $\bar{y}^{1/2}=\bar{\gamma}\sqrt{1-\frac{4J^2c^2}{r^2E^2}}$ with this approximation.

\section{Lagrangian formalism in harmonic coordinates}
\label{sec:lag}

\subsection{The general case}
\label{sec:lagA}

In this section we look for a Lagrangian associated with our general equations of motion, for any $N$, in harmonic coordinates, see Eqs.~\eqref{eom1}--\eqref{eom2}. The Lagrangian will be given by the special-relativistic expression plus terms of order $G$, and we neglect higher-order terms in $G$. It is known that the Lagrangian in harmonic coordinates is a generalized one, depending not only on positions and velocities $\bm{y}_a, \bm{v}_a$ but also on the accelerations $\bm{a}_a=\ud\bm{v}_a/\ud t$ of the particles. In a PN expansion the accelerations appear at order 2PN~\cite{DD81b,DS85} and they are contained in terms linear in $G$, see Eq.~(209) in~\cite{BlanchetLR}. Of course, replacing the accelerations by the equations of motion would yield negligible terms of order $G^2$. However, it is not allowed to replace accelerations in a Lagrangian while remaining in the same coordinate system. Such replacement is equivalent to a shift in the particles' trajectories (or ``contact'' transformation), \textit{i.e.} the new Lagrangian is physically equivalent to the original one but written in a different coordinate system~\cite{S84}. We shall confirm this result within our 1PM framework. Furthermore, by employing the technique of double-zero (or multiple-zero) terms~\cite{DS85}, it is sufficient to consider a Lagrangian that is \textit{linear} in accelerations. Indeed, the procedure can work for any PN order, and is thus formally valid at the 1PM order. Therefore, we look for the Lagrangian in the form 
\begin{equation}\label{Ldef}
L[y, v, a] = - \sum_a \frac{m_a c^2}{\gamma_a} + \lambda + \sum_a q_a^i a_a^i\,,
\end{equation}
where we symbolize the functional dependence of $L$ by $L[y, v, a] \equiv L[\{\bm{y}_a, \bm{v}_a, \bm{a}_a\}]$, where $\lambda$ and $q_a^i$ are of order $G$ and both depend only on positions and velocities, \textit{i.e.} $\lambda[y, v]$ and $q_a^i[y, v]$. The linear dependence in accelerations is made explicit in~\eqref{Ldef}. Denoting $p_a^i$ and $q_a^i$ the conjugate momenta associated with the positions $y_a^i$ and velocities $v_a^i$, \textit{i.e.}
\begin{subequations}\label{paqa}\begin{align}
p_a^i =& \frac{\delta L}{\delta v_a^i} = \frac{\partial
L}{\partial v_a^i}-\frac{\ud}{\ud t} \bigg( \frac{\partial L}{\partial
a_a^i} \bigg)\,, \label{pa}\\ q_a^i =& \frac{\delta L}{\delta
a_a^i} = \frac{\partial L}{\partial a_a^i}\,,\label{qa}
\end{align}\end{subequations}
the equations of motion take the form
\begin{equation}\label{eomfull}
\frac{\ud p_a^i}{\ud t} = \frac{\partial L}{\partial y_a^i}\,.
\end{equation}
The conserved energy is given by the generalized Legendre transformation (see \textit{e.g.}~\cite{ABF01,Wood07})
\begin{equation}\label{conservedE}
E = \sum_a \Bigl( p_a^i v_a^i + q_a^i a_a^i \Bigr) - L\,.
\end{equation}
In both Eqs.~\eqref{eomfull} and~\eqref{conservedE} the accelerations have to be replaced by the equations of motion. For instance, the term $q_a^i a_a^i$ in $E$ is second-order in $G$ and can be neglected at 1PM order (but note that it cancels anyway with the same term in $L$). Using the form of the Lagrangian~\eqref{Ldef} we obtain the equations of motion as
\begin{equation}\label{eomexpl}
f_a^i = \frac{\delta \lambda}{\delta y_a^i} + \ddot{q}_a^i \,.
\end{equation}
Here we define the force $f_a^i= m_a \frac{\ud}{\ud t}(\gamma_a v_a^i)$ and $\frac{\delta \lambda}{\delta y_a^i} = \frac{\partial
\lambda}{\partial y_a^i}-\frac{\ud}{\ud t} ( \frac{\partial \lambda}{\partial
v_a^i} )$, and the dots refer to time derivatives.\footnote{We have $f^i=4f_1^i=-4f_2^i$ in the notation of Eq.~\eqref{force}.} On the other hand, the integral of the energy is given by~\eqref{Edef} with
\begin{equation}\label{eqV}
V = \sum_a v_a^i \frac{\partial \lambda}{\partial v_a^i} - \lambda - \sum_a v_a^i \dot{q}_a^i \,,
\end{equation}
The left-hand sides (LHS) of Eqs.~\eqref{eomexpl} and~\eqref{eqV} are known, since we have determined $f_a^i$ and $V$ in~\eqref{eom1}--\eqref{eom2} and~\eqref{V}. However, the two equations are not independent, since $f_a^i$ and $V$ satisfy the constraint
\begin{equation}\label{constraint}
\frac{\ud V}{\ud t} + \sum_a v_a^i f_a^i = 0\,.
\end{equation}

Our strategy is to first find a particular solution, say $(\hat{\lambda}, \hat{q}_a^i)$, such that the set of conjugate momenta $\hat{q}_a^i$ obey the equation
\begin{equation}\label{equation}
\sum_a v_a^i\hat{q}_a^i=0 \,.
\end{equation}
To order $G$, the same constraint is also satisfied by the time derivative $\dot{\hat{q}}_a^i$, therefore, for the solution $(\hat{\lambda}, \hat{q}_a^i)$, the equation~\eqref{eqV} reduces to an ordinary Legendre transformation,
\begin{equation}\label{PDE}
V = \sum_a v_a^i \frac{\partial \hat{\lambda}}{\partial v_a^i} - \hat{\lambda} \,.
\end{equation}
In order to determine $\hat{\lambda}$, we note that the potential $V$ given in~\eqref{V} reduces in the limit $c\to +\infty$ to the Newtonian approximation, namely $V = U + \mathcal{O}(1/c^2)$ where the Newtonian potential is
\begin{equation}\label{VN}
U = - \sum_{a<b} \frac{G m_a m_b}{r_{ab}} \,.
\end{equation}
If we subtract the Newtonian limit $U$, we get a quantity which tends to zero when $c\to +\infty$ like $\mathcal{O}(1/c^2)$. Then, it is straightforward to show that a well-behaved solution of Eq.~\eqref{PDE} is
\begin{equation}\label{lambdasol}
\hat{\lambda} = - U + \frac{1}{c}\int_{c}^{+\infty} \ud s \biggl[ V\Bigl(\bm{y}_b, \frac{\bm{v}_b}{s}\Bigr) - U(\bm{y}_b)\biggr]\,.
\end{equation}
Indeed, we insert into Eq.~\eqref{V} all the relevant factors $c$ and make the replacement $c\longrightarrow s$, and then integrate over the ``speed of light'' $s$ from the physical value $c$ up to infinity. The bound $s\to\infty$ corresponds to the Newtonian limit and we see from the definition of the Newtonian potential~\eqref{VN} that the integral is convergent. The first term in Eq.~\eqref{lambdasol} represents the Newtonian approximation with the correct minus sign for a Lagrangian, and the integral represents formally the complete series of PN corrections, but resummed in the PM approximation. The result~\eqref{lambdasol} can be rewritten in a simpler way as the ``partie finie'' (PF) of the integral in the sense of Hadamard~\cite{Hadamard}, taking care of the divergence at infinity:
\begin{equation}\label{lambdasolHad}
\hat{\lambda} = \text{PF} \,\frac{1}{c}\int_{c}^{+\infty} \ud s \,V\Bigl(\bm{y}_b, \frac{\bm{v}_b}{s}\Bigr) \,.
\end{equation}
Note that for this very simple type of divergence $\sim s^0 + \mathcal{O}(s^{-2})$ the PF does not depend on an arbitrary constant. Unfortunately, given the complicated structure of $V$ in Eq.~\eqref{V}, we have not been able to perform explicitly the integration and obtain a closed form expression in the general case. But we shall discuss in Sec.~\ref{sec:eqmass} the equal mass case for which an analytical form exists. 

Equations~\eqref{lambdasol} or~\eqref{lambdasolHad} give a solution of the equation~\eqref{eqV} but we still have to adjust $\hat{q}_a^i$ in order to satisfy the equations of motion~\eqref{eomexpl}. Thus, we look for $\hat{q}_a^i$ satisfying 
\begin{equation}\label{eomexpl2}
\ddot{\hat{q}}_a^i = f_a^i - \frac{\delta \hat{\lambda}}{\delta y_a^i} \,,
\end{equation}
where the RHS is now known. Since it is of order $G$ we can integrate following the same method as used in Sec.~\ref{sec:consE}. That is, we transform~\eqref{eomexpl2} into an ordinary differential equation using the change of variable $t \longrightarrow r_{ab}$, and then integrate twice to determine $\dot{\hat{q}}_a^i$, then $\hat{q}_a^i$. The solution automatically satisfies the constraint~\eqref{equation} by virtue of~\eqref{constraint}.

Finally we have found a particular Lagrangian associated with our equations of motion, in the form of the particular solution $(\hat{\lambda}, \hat{q}_a^i)$. Now the general solution $(\lambda, q_a^i)$ can be obtained by adding an arbitrary total time-derivative $\ud F/\ud t$, where $F$ is a function of the positions $y_a^i$ and velocities $v_a^i$; indeed, by definition of the functional derivative, $\delta(\ud F/\ud t)/\delta y_a^i=0$. Hence the general solution (for the class of harmonic-coordinate Lagrangians that are linear in accelerations) reads
\begin{subequations}\label{gaugefreedom}
\begin{align}
\lambda &= \hat{\lambda} + \sum_a v_a^i \frac{\partial F}{\partial y_a^i} \,,\\
q_a^i &= \hat{q}_a^i + \frac{\partial F}{\partial v_a^i} \,.
\end{align}\end{subequations}
%

At the 1PM order we find that the Lagrangian in harmonic coordinates irreducibly depends on accelerations, \textit{i.e.}, it is impossible to determine $F$ such that $q_a^i=0$. However, we know that the accelerations in a Lagrangian can always be eliminated by appropriate shifts of the trajectories. In the present case the shifts are given by $Y_a^i = y_a^i + \xi_a^i$, where $\xi_a^i$ is of order $G$ and is determined by the conjugate momenta $q_a^i$. We find (adapting techniques from Refs.~\cite{S84,DS85,DS91,ABF01})
\begin{equation}\label{shift}
\xi_a^i = -\frac{1}{m_a \gamma_a} \left[ q_a^i - \frac{v_a^i (v_a q_a)}{c^2}\right]\,.
\end{equation}
Then, an \textit{ordinary} Lagrangian, dynamically equivalent to the harmonic-coordinate one~\eqref{Ldef}, but valid for the shifted trajectories $Y_a^i$ instead, is given to order $G$ by
\begin{equation}\label{L'}
L'[Y, V] = - \sum_a \frac{m_a c^2}{\gamma_a} + \lambda \,,
\end{equation}
where $\lambda$ and $\gamma_a$ denote the same functionals as before, but expressed in terms of the shifted variables, \textit{i.e.}, $\lambda[Y,V]$ and $\gamma_a[V]$. The conserved energy is given by $E=\sum_a m_a c^2 \gamma_a + V'$ but with the modified potential function
\begin{equation}\label{V'}
V' = V + \sum_a V_a^i q_a^i \,.
\end{equation}
Hence, in this construction, the particular solution $\hat{\lambda}$ found in Eqs.~\eqref{lambdasol}--\eqref{lambdasolHad} represents the equivalent ordinary Lagrangian in shifted variables for which the energy functional remains the same ($V'=V$); compare~\eqref{equation} and~\eqref{V'}.

In spite of the fact that it has not been possible to find a closed-form expression for the integral~\eqref{lambdasol} in the general case, the relevant integral can easily be worked out in the PN limit $c\to\infty$. We start from the known 4PN expansion of the potential $V$ and explicitly perform the integration~\eqref{lambdasol} term by term, to obtain the corresponding 4PN expansion of $\hat{\lambda}$. Then, we derive the 4PN result for $\hat{q}_a^i$ by the method indicated previously. Finally, we find a unique total time-derivative, with some function $F_\text{PN}$ given in the form of a PN expansion, so that the Lagrangian agrees up to order $G$ with the 4PN Lagrangian published in Refs.~\cite{BBBFMc,BBFM17}. In conclusion, the 4PN limit of our Lagrangian is correct. The function $F_\text{PN}$ is given by Eqs.~\eqref{functionF}--\eqref{FPN} in Appendix~\ref{app:5PN}. Furthermore we have extended the analysis to higher order and give the terms of order $G$ in the harmonic coordinates Lagrangian up to 5PN order in Appendix~\ref{app:5PN}, see~\eqref{lambdaPN}--\eqref{q1PN} there.

\subsection{The two-body equal-mass case}
\label{sec:lagB}

We apply the previous procedure for two bodies ($N=2$) with equal masses, following the notation~\eqref{eqmassnot}. The harmonic coordinates Lagrangian becomes in this case
\begin{equation}\label{Leqmass}
L = - \frac{m c^2}{\gamma} + \lambda + q^i a^i\,,
\end{equation}
where $q^i=\frac{1}{2}(q_1^i-q_2^i)$ is the conjugate momentum of the relative velocity $v^i$, as defined by $q^i = \partial L/\partial a^i$. We follow the method of Sec.~\ref{sec:lagA} to determine $\lambda$ and $q^i$. We first look for a particular solution $(\hat{\lambda}, \hat{q}^i)$ such that $v^i\hat{q}^i=0$. Then, we determine $\hat{\lambda}$ from the potential function $V$ determined for the case at hands in Sec.~\ref{sec:eqmass} as
\begin{equation}\label{Veqmass}
V = \frac{G m^2}{4 r y^{1/2}} \left(2\gamma^2-1\right)\Bigl(8\gamma^3-8\gamma-\frac{1}{\gamma}\Bigr)\,.
\end{equation}
Substituting~\eqref{Veqmass} into Eq.~\eqref{lambdasolHad} we can integrate and obtain a rather complicated result:
\begin{align}\label{lambdahat}
\hat{\lambda} &= \frac{G m^2}{4 r \gamma}\left(1+\frac{8(\gamma^2-1)^2}{y-1}\right)\sqrt{y} + \frac{G m^2}{4 r}\frac{\gamma^2-1}{\gamma\sqrt{\gamma^2-y}}\left( \arctan\sqrt{\frac{y}{\gamma^2-y}} - \frac{\pi}{2}\right) \nonumber\\& \qquad\qquad\qquad - \frac{2G m^2}{r}\frac{(\gamma^2-1)(\gamma^2-y)}{\gamma(y-1)^{3/2}} \ln\left( \sqrt{y} + \sqrt{y-1} \right)\,.
\end{align}
Note that the circular orbit limit of this expression, $(nv)\to 0$ or $y\to 1$, is well defined, as can be verified directly on Eq.~\eqref{lambdahat}.

As discussed in Sec.~\ref{sec:lagA}, the Lagrangian $\hat{L} = - \frac{m c^2}{\gamma} + \hat{\lambda}$ represents an equivalent Lagrangian which is ordinary (no accelerations), and valid in a coordinate system shifted with respect to the harmonic-coordinate system, with shift $\xi^i = -\frac{4}{m \gamma}\,\hat{q}^i$, see Eq.~\eqref{shift}. We shall investigate in Sec.~\ref{sec:Heqmass} the connection of this Lagrangian to the 1PM Hamiltonian of~\cite{LSB08}.

Next, we have to determine $\hat{q}^i$ using the method of Sec.~\ref{sec:lagA}. The computation is straightforward and we obtain an expression which contains transcendental functions like in~\eqref{lambdahat}. We do not give the complete expression because we can now adjust the arbitrary function $F$ in Eqs.~\eqref{gaugefreedom} in order to simplify the result as much as possible. In fact we find that all the transcendental functions ($\arctan x$ and $\ln x$) in both $\hat{\lambda}$ and $\hat{q}^i$ can be removed by the following choice:
\begin{align}\label{F}
F &= - \frac{G m^2}{8}\sqrt{\frac{y-1}{\gamma^2-y}}\left( \arctan\sqrt{\frac{y}{\gamma^2-y}} - \frac{\pi}{2}\right) \nonumber\\& \qquad\qquad\qquad - \frac{G m^2}{2}\frac{\gamma^2-y}{y-1} \Bigl[ \ln\left( \sqrt{y} + \sqrt{y-1} \right) - \sqrt{y-1} \Bigr]\,.
\end{align}
In the second term, we have subtracted from the logarithm the term $\sqrt{y-1}$ in order to ensure the well-defined circular orbit limit $y \to 1$. Finally, adding up the total time derivative $\ud F/\ud t$ [see Eqs.~\eqref{gaugefreedom}] we obtain the expressions: 
\begin{subequations}\label{lambdaqi}
\begin{align}
\lambda &= \frac{G m^2}{4 r}\frac{4 y^2-8y+1 + \sqrt{y}\bigr(8 \gamma^4-8\gamma^2+1\bigr) + 4\gamma^4}{\gamma(y + \sqrt{y})}\,,\\
q^i &= \frac{G m^2}{16} \biggl[\frac{4 \gamma^4 - 4 y \gamma^2 +1 + \sqrt{y}\bigr(8 \gamma^4-8\gamma^2+1\bigr)}{\gamma(y + \sqrt{y})} \,n^i \nonumber\\ 
& \qquad\qquad + \frac{4 \gamma^4 - 4 y \gamma^2 +1 - \sqrt{y}\bigr(8\gamma^2-1\bigr)}{2(y + \sqrt{y})} \sqrt{y-1}\,v^i \biggr] \,,
\end{align}\end{subequations}
which completely specify our acceleration dependent Lagrangian~\eqref{Ldef}. By varying this Lagrangian we can recover the equations of motion~\eqref{singleeom}--\eqref{force}.

\section{Hamiltonian formalism}
\label{sec:Ham}

\subsection{Equal-mass case}
\label{sec:Heqmass}

Let us rewrite in full form the ordinary Lagrangian we obtained in the previous section, $\hat{L} = - \frac{m c^2}{\gamma} + \hat{\lambda}$, and which we recall is not valid in harmonic coordinates but in a shifted coordinate system $X^i = x^i + \xi^i$, see Eq.~\eqref{shift}. Using the more common notations $V^2$ and $(NV)=\dot{R}$ (with $c=1$ and removing the ``hat''), we find
\begin{align}\label{Lhat}
L[X,V] &= - m \sqrt{1-\frac{V^2}{4}} + \frac{G m^2}{8R}\left(1+\frac{8V^4}{(NV)^2(4-V^2)}\right)\sqrt{4+(NV)^2-V^2} \nonumber\\& \qquad\qquad\quad + \frac{G m^2}{8 R}\frac{V^2}{\sqrt{V^2-(NV)^2}}\left( \arctan\sqrt{\frac{4+(NV)^2-V^2}{V^2-(NV)^2}} - \frac{\pi}{2}\right) \nonumber\\& \qquad\qquad\quad - \frac{G m^2}{R}\frac{V^2(V^2-(NV)^2)}{(NV)^3} \ln\left(\frac{(NV)+\sqrt{4+(NV)^2-V^2}}{\sqrt{4-V^2}} \right)\,.
\end{align}
The Hamiltonian is easily constructed by an ordinary Legendre transformation, and we obtain, up to order $G$ (posing $\overline{m}=\sqrt{m^2+4P^2}$)
\begin{align}\label{Hhat}
H[X,P] &= \overline{m} - \frac{G}{4R\,\overline{m}}\left(m^2+\frac{32P^4}{(NP)^2}\right)\sqrt{m^2+4(NP)^2} \nonumber\\& \qquad\quad - \frac{G m^2}{2R\,\overline{m}}\frac{P^2}{\sqrt{P^2-(NP)^2}}\left( \arctan\sqrt{\frac{m^2+4(NP)^2}{4(P^2-(NP)^2)}} - \frac{\pi}{2}\right) \nonumber\\& \qquad\quad + \frac{4G m^2}{R\,\overline{m}}\frac{P^2(P^2-(NP)^2)}{(NP)^3} \ln\left(\frac{2(NP)+\sqrt{m^2+4(NP)^2}}{m} \right)\,.
\end{align}
It is very interesting to compare this Hamiltonian with the one obtained by Ledvinka, Sch{\"a}fer and Bi\v{c}ak (LSB)~\cite{LSB08}. These authors obtained a nice, closed-form expression for the Hamiltonian in the general case, see Eq.~(11) in Ref.~\cite{LSB08}. On the other hand, the specific method explained in Sec.~\ref{sec:lag} did not lead to a closed form expression for the Lagrangian except in the equal mass case, yielding the results~\eqref{Lhat}--\eqref{Hhat}.

We want to find a canonical transformation between our Hamiltonian~\eqref{Hhat} and the LSB one~\cite{LSB08}. In particular, this canonical transformation should be able to remove the transcendental functions ($\arctan x$ and $\ln x$) from the Hamiltonian~\eqref{Hhat}, since there are no such functions in the LSB Hamiltonian. This is easy to do, as we have already seen in Eq.~\eqref{F}, where the transcendental functions at the level of the Lagrangian can be removed by adding an appropriate total time-derivative to the Lagrangian. 

To find the canonical transformation we adopt reduced canonical variables $(R, \Phi, P_R, P_\Phi)$ by posing $P^2=P_R^2+P_\Phi^2/R^2$ and $P_R=(NP)$. Then, we look for a canonical transformation  
\begin{equation}\label{cantransf}
(R, \Phi, P_R, P_\Phi) \longrightarrow (R', \Phi', P_R', P_\Phi')\,,
\end{equation}
associated with some generating function $M^\text{gen}(R, \Phi, P_R', P_\Phi')$. We assume no explicit time dependence of the generating function so that
\begin{equation}\label{HH'}
H(R, \Phi, P_R, P_\Phi)=H'(R', \Phi', P_R', P_\Phi')\,.
\end{equation}
Furthermore, since the special-relativistic limits of both Hamiltonians agree, the generating function should be of the type
\begin{equation}\label{generating}
M^\text{gen} = R\,P_R' + \Phi\, P_\Phi' +  F^\text{gen}(R, P_R', P_\Phi') \,,
\end{equation}
where the function $F^\text{gen}(R, P_R', P_\Phi')$ is of order $G$. Finally, both Hamiltonians are rotationally invariant, so the function $F^\text{gen}$ should not depend on $\Phi$. To order $G$ the transformation laws between the two sets of variables are as follows:
\begin{subequations}\label{translaw}
\begin{align}
R' &= R + \frac{\partial F^\text{gen}}{\partial P_R}(R, P_R, P_\Phi)\,,\\
\Phi' &= \Phi + \frac{\partial F^\text{gen}}{\partial P_\Phi}(R, P_R, P_\Phi)\,,\\
P_R' &= P_R - \frac{\partial F^\text{gen}}{\partial R}(R, P_R, P_\Phi)\,,
\end{align}
\end{subequations}
and, in addition, $P_\Phi' = P_\Phi$, which implies that the conserved angular momentum of both Hamiltonians is the same to order $G$. Inserting Eqs.~\eqref{translaw} into~\eqref{HH'} we obtain (to order $G$)
\begin{equation}\label{HH'expl}
H'(R, \Phi, P_R, P_\Phi) = H(R, \Phi, P_R, P_\Phi) - \frac{4}{\overline{m}}\left( P_R \frac{\partial F^\text{gen}}{\partial R} + \frac{P_\Phi^2}{R^3} \frac{\partial F^\text{gen}}{\partial P_R}\right)\,.
\end{equation}

We proceed in two steps. We first apply a generating function in order to remove the transcendental functions from Eq.~\eqref{Hhat}. As the effect of a canonical transformation with generating function $F^\text{gen}$ is the same as the effect of adding a total time derivative in the Lagrangian formalism, we may use for the generating function the one already computed in Eq.~\eqref{F}, hence $F^\text{gen} = F$, with
\begin{align}\label{fgener}
F &= - \frac{G m^2 R P_R}{8P_\Phi} \left( \arctan\left(\frac{R\sqrt{m^2 + 4P_R^2}}{2P_\Phi}\right) - \frac{\pi}{2}\right) \nonumber\\& \qquad\qquad\qquad - \frac{G m^2 P_\Phi^2}{2R^2 P_R^2} \left[ \ln\left( \frac{2P_R + \sqrt{m^2 + 4P_R^2}}{m} \right) - \frac{2P_R}{m} \right]\,.
\end{align}
Under this canonical transformation our Hamiltonian~\eqref{Hhat} is very much simplified. Posing $y = 1 + 4(NP)^2/m^2$ (consistently with our notation in previous sections) we get
\begin{equation}\label{H'expl}
H' = \overline{m} + \frac{4 G m}{R\,\overline{m}}\left(\frac{P^4-(NP)^4}{(NP)^2} + \frac{1}{\sqrt{y}}\left[-\frac{m^2}{16}+(NP)^2-2P^2-8\frac{P^4}{m^2}-\frac{P^4}{(NP)^2}\right]\right)\,.
\end{equation}
On the other hand, the LSB Hamiltonian, Eq.~(11) in~\cite{LSB08} for two equal masses, is
\begin{equation}\label{HLSB}
H_\text{LSB} = \overline{m} + \frac{G m}{R}\left( - 7 P^2 - (NP)^2 + \frac{m^3}{4\overline{m}\,\sqrt{y}}\right)\,,
\end{equation}
so it must be a simple task to find the canonical transformation linking the two Hamiltonians. Indeed, we find that there exists such a canonical tranformation, but which is not so simple as it does not admit a closed form expression, involving the two-variable hypergeometric Appell function $F_1(\alpha;\beta,\beta';\gamma;x,y)$~\cite{AbramStegun}. We finally get
\begin{align}\label{deltaFLSB}
\delta F_\text{LSB} =& -\frac{G \overline{m} P_R}{4} -\frac{G m P_\Phi^2}{R^2 P_R}  + \frac{G m}{\sqrt{y}}\left(P_R + \frac{4 P_R^3}{m^2} + \frac{P_\Phi^2}{R^2 P_R} + \frac{4 P_R\,P_\Phi^2}{m^2\,R^2}\right) \nonumber\\ & -G P \left(2\overline{m}+\frac{m^2}{\overline{m}}\right)\ln\Bigl(R\left[\overline{m} P_R+m\,P\,\sqrt{y}\right]\Bigr) + 2 G \overline{m}\,P\,\ln\Bigl(R\,P\left[P + P_R\right]\Bigr)\nonumber\\ & + \frac{G m\,P\,P_R^3R^3}{3 P_\Phi^3} \,F_1\left(\tfrac{3}{2};\tfrac{1}{2},1;\tfrac{5}{2};-\tfrac{\overline{m}^2P_R^2R^2}{m^2P_\Phi^2},-\tfrac{P_R^2R^2}{P_\Phi^2}\right)\,.
\end{align}
Hence, the complete canonical transformation linking our Hamiltonian to the LSB one in the equal mass case is generated by $F + \delta F_\text{LSB}$. However, as we said we have not been able to connect the two Hamiltonians in the more general case.

\subsection{Frame of the center of mass}
\label{sec:CM}

In special relativity the linear momentum of non interacting particles is $m_a\gamma_a v_a^i$, and we have obtained in Sec.~\ref{sec:consE} the total linear momentum at the 1PM order (in the case of two particles with arbitrary masses):
\begin{equation}\label{Ptotal}
P^i =  m_1\gamma_1 v_1^i + m_2\gamma_2 v_2^i + \Pi^i\,,
\end{equation}
where $\Pi^i$ is the correction of order $G$ given by Eq.~\eqref{Pi}.

The frame of the center of mass (CM) is defined by $\bm{P}=0$. In this frame the dynamics will be described by dynamical variables $(\bm{x}, \bm{v})$ where $\bm{x}=\bm{y}_1-\bm{y}_2$ and $\bm{v}=\bm{v}_1-\bm{v}_2$ are the relative variables. The CM Lagrangian in harmonic coordinates will be a functional of $\bm{x}$, $\bm{v}$ and also the acceleration $\bm{a}$. In the Hamiltonian formalism the CM dynamics is described by canonical variables $(\bm{x}, \bm{p})$ where $\bm{p}$ is the CM linear momentum $\bm{p}=\bm{p}_1=-\bm{p}_2$. In special relativity we have $\bm{v}_1=\bm{p}/\overline{m}_1$ and $\bm{v}_2=-\bm{p}/\overline{m}_2$, where we recall that $\overline{m}_a=\sqrt{m_a^2+p^2}$ (and pose $c=1$). At the 1PM order, let us pose with full generality, for the individual velocities,
\begin{subequations}\label{v12CM}
\begin{align}
v_1^i &=  \frac{p^i + Y^i}{\overline{m}_1} + W^i\,, \\
v_2^i &=  - \frac{p^i + Y^i}{\overline{m}_2} + W^i\,,
\end{align}
\end{subequations}
where $Y^i$ and $W^i$ represent the 1PM corrections of order $G$, with $Y^i$ (like $p^i$) changing sign when we exchange the two particles' labels, and with $W^i$ staying invariant under such exchange $1\leftrightarrow 2$. The relative velocity is given by
\begin{equation}\label{vCM}
v^i = \frac{p^i + Y^i}{\overline{\mu}}\,,
\end{equation}
where the associated reduced mass is $\overline{\mu}=\overline{m}_1\overline{m}_2/\overline{m}$ where $\overline{m}=\overline{m}_1+\overline{m}_2$ is the total mass. From Eqs.~\eqref{v12CM} one readily derives (to order $G$)
\begin{subequations}\label{relCM}
\begin{align}
m_1 \gamma_1 v_1^i &=  p^i + X^i + \overline{m}_1 \left( W^i + \frac{p^i}{m_1^2}(pW)\right) \,, \\ 
m_2 \gamma_2 v_2^i &=  - p^i - X^i + \overline{m}_2 \left( W^i + \frac{p^i}{m_2^2}(pW)\right) \,,
\end{align}
\end{subequations}
where $(pW)$ denotes the scalat product as usual, and where we introduce the intermediate notation $X^i$ defined by the (easily invertible) relation\footnote{Denoting $\overline{\nu}=\overline{\mu}/\overline{m}$ we have $$\frac{\partial \ln\overline{\mu}}{\partial p^i} = \frac{1 - 3\overline{\nu}}{\overline{\nu}^2}\frac{p^i}{\overline{m}}\,.$$}
\begin{equation}\label{intermediate}
Y^i = X^i - (pX) \frac{\partial \ln\overline{\mu}}{\partial p^i}\,.
\end{equation}
The expressions~\eqref{relCM} are ready for insersion into the total linear momentum~\eqref{Ptotal} and the CM prescription $P^i=0$. Although $X^i$ is clearly left undetermined at this stage, the CM prescription specifies $W^i$ in terms of the known 1PM correction in the linear momentum, given by~\eqref{Pi}, and we find
\begin{equation}\label{Wi}
W^i = - \frac{1}{\overline{m}_1+\overline{m}_2}\left[ \Pi^i - \frac{\overline{m}_1 m_2^2+\overline{m}_2 m_1^2}{\overline{m}^3_1 m_2^2+\overline{m}^3_2 m_1^2} \,(p\Pi) \,p^i\right]\,.
\end{equation}
The CM Hamiltonian is then obtained starting from the expression of the energy $E=m_1\gamma_1+m_2\gamma_2+V$, where $V$ is given by~\eqref{V}, by replacing the velocities $v_1^i$, $v_2^i$ in terms of the CM canonical momentum $p^i$ using~\eqref{v12CM}. In this way, we obtain the CM Hamiltonian to order $G$ in the form 
\begin{equation}\label{H0}
H = \overline{m}_1+\overline{m}_2 + \left(\frac{1}{\overline{m}_1}+\frac{1}{\overline{m}_2}\right)(pX) + K\,,
\end{equation}
where $(pX)$ is still undetermined, but $K$ is known at this stage:
\begin{equation}\label{K}
K = - \frac{\overline{m}_1^2 m_2^2-\overline{m}_2^2 m_1^2}{\overline{m}^3_1 m_2^2+\overline{m}^3_2 m_1^2} \,(p\Pi) + V\,,
\end{equation}
or, after an explicit calculation,
\begin{align}\label{Kexpl}
K &= - \frac{\overline{m}_1 \overline{m}_2^2+p^2}{\overline{m}^3_1 m_2^2+\overline{m}^3_2 m_1^2}\Bigl(\overline{m}^2_1\overline{m}^2_2 - \left(3\overline{m}^2_1 + 4 \overline{m}_1 \overline{m}_2 + 3\overline{m}^2_2\right)p^2 + p^4\Bigr)\nonumber\\&\qquad\qquad\qquad\qquad
\times\left[ \frac{G \overline{m}^2_1}{r \sqrt{m_1^2 + (np)^2}} + \frac{G \overline{m}^2_2}{r \sqrt{m_2^2 + (np)^2}}\right]\,.
\end{align}

Now with the previous form~\eqref{H0} of the Hamiltonian, we can determine $X^i$ (or equivalently $Y^i$), recalling the definition~\eqref{vCM} of the velocity, by using the Hamiltonian equation $v^i=\partial H/\partial p^i$ which is easily seen to be equivalent to the equation
\begin{equation}\label{Hameq}
p^i\,\frac{\partial X^j}{\partial p^i} + \overline{\mu}\,\frac{\partial K}{\partial p^i} = 0\,.
\end{equation}
The latter equation, after contraction with $p^i$, implies in turn the partial differential equation
\begin{equation}
p^i\,\frac{\partial (pX)}{\partial p^i} - (pX) = - \overline{\mu}\,p^i\,\frac{\partial K}{\partial p^i}\,,
\end{equation}
which can in principle be solved to obtain $(pX)$ and the Hamiltonian~\eqref{H0}. However the source term in the RHS involves the partial derivatives of the expression~\eqref{Kexpl}, which are rather cumbersome to compute in practice, so we prefer solving the equivalent equation obeyed by the auxiliary quantity $\sigma=(pX)+\overline{\mu}\,K$, which is
\begin{equation}
p^i\,\frac{\partial \sigma}{\partial p^i} - \sigma = - \overline{\mu}\,K\left( 1 - p^i\,\frac{\partial \ln \overline{\mu}}{\partial p^i}\right)\,,
\end{equation}
and whose source term $S=- \overline{\mu}\,K ( 1 - p^i\,\frac{\partial \ln \overline{\mu}}{\partial p^i})$ is simpler. Not surprisingly, the latter equation has the same form as the one found in the Lagrangian formalism, Eq.~\eqref{PDE}, and we solve it by means of the same technique, \textit{i.e.} (with PF denoting the Hadamard partie finie)
\begin{equation}\label{sigmasolHad}
\sigma = \text{PF} \,\frac{1}{c}\int_{c}^{+\infty} \ud s \,S\Bigl(\bm{x}, \frac{\bm{p}}{s}\Bigr) \,.
\end{equation}
Like for the Lagrangian formalism we could not obtain a closed form expression for the integral~\eqref{sigmasolHad} in the general case, but only in the equal mass case $m_1=m_2$, in which case we simply recover the Hamiltonian~\eqref{Hhat}.

\acknowledgments

One of us (L.B.) would like to thank Gilles Esposito-Far\`ese for interesting discussions and for comments on a preliminary version of the manuscript.

\appendix

\section{Gravitational scattering of two particles}
\label{app:scatt}

We compute the scattering of two particles with arbitrary masses using the formalism developed in Secs.~\ref{sec:eom} and~\ref{sec:consE}. The total change of linear momentum of the particle 1 between the ``in'' and ``out'' states at $t=\pm \infty$ is given by
\begin{equation}\label{Deltap1}
\Delta p_1^i = m_1 c \Delta u_1^i = m_1 c \int_{-\infty}^{+\infty} \ud\tau_1\frac{\ud u_1^i}{\ud\tau_1}\,.
\end{equation}
The equations of motion of the particle 1 in the field of particle 2 are given by Eq.~\eqref{eom1}, therefore the total change in the spatial components of the linear momentum is given by the equation
\begin{align}\label{changep1in}
\Delta p_1^i &= - G m_1 m_2 \,\gamma_1^{-1}\int_{-\infty}^{+\infty}\frac{\ud t}{r_{12}^2 \,y_{12}^{3/2}}\biggl[ (2\epsilon_{12}^2-1)n_{12}^i \\ & \qquad\quad + (2\epsilon_{12}^2+1)\Bigl( - (n_{12} u_1) + \epsilon_{12} (n_{12} u_2)\Bigr) u_1^i + \Bigl( 4 \epsilon_{12} (n_{12} u_1) - (2\epsilon_{12}^2+1) (n_{12} u_2) \Bigr) u_2^i\biggr] \,,\nonumber
\end{align}
whereas the total change in the zero-th component of the momentum is $\Delta p_1^0=\frac{v_1^i}{c}\Delta p_1^i$. 

Recall that to order $G$ we can integrate the motion on straight lines, therefore all velocity dependent factors in the integrand of~\eqref{changep1in} are in fact constant, except those involving the separation $r_{12}$ and unit direction $n_{12}^i$. Therefore, we see immediately that the integral~\eqref{changep1in} can be expressed by means of the solution $D^i_{12}$ of the elementary equation~\eqref{dDab}. Denoting the total change in this quantity by $\Delta D^i_{12} = D^i_{12}(+\infty) - D^i_{12}(-\infty)$, we obtain
\begin{align}\label{changep1}
\Delta p_1^i = - G m_1 m_2 \,\gamma_1^{-1} \biggl[ & (2\epsilon_{12}^2-1)\Delta D^i_{12} + (2\epsilon_{12}^2+1)\Bigl( - (\Delta D_{12} u_1) + \epsilon_{12} (\Delta D_{12} u_2)\Bigr) u_1^i \nonumber\\ & \qquad\quad + \Bigl( 4 \epsilon_{12} (\Delta D_{12} u_1) - (2\epsilon_{12}^2+1) (\Delta D_{12} u_2) \Bigr) u_2^i\biggr] \,,
\end{align}
where we denote, for instance, $(u_1\Delta D_{12})=\Delta D^i_{12} u_1^i = \gamma_1 \Delta \bm{D}_{12}\cdot\bm{v}_1/c$.

The explicit expression of the solution $D_{12}^i$ has been obtained in Eq.~\eqref{closedform}; again, since the velocities are constant, the only time dependent variable in this expression is the separation $r_{12}$. At $t=+\infty$ in the future we have 
$r_{12}\to +\infty$ and we readily find the limit from Eq.~\eqref{closedform}. When $t=-\infty$ we must choose the part of the trajectory for which $r_{12}$ decreases rather than increases, and the limit just changes sign, adding finally a factor 2. Thus, the total change in this quantity is given by
\begin{equation}\label{DeltaD0}
\Delta D^i_{12} = \frac{2}{C_{12}^2+\frac{\gamma_2^2}{c^2}\left[(\mu_{12}v_2)^2v_{12}^2+C_{12}^2\frac{(v_{12}v_2)^2}{v_{12}^2}\right]}\frac{\mu^i_{12}\vert v_{12}\vert}{\sqrt{1+\frac{\gamma_2^2}{c^2}\frac{(v_{12}v_2)^2}{v_{12}^4}}} \,.
\end{equation}
We then return to the original variables, as we did in~\eqref{Dabsol}. Since the result~\eqref{DeltaD0} is constant, it can be evaluated at any point along the trajectory. We choose to evaluate it at the minimal distance of approach of the two particles, characterized by $\dot{r}_{12}=(n_{12}v_{12})=0$. At this point, $r_{12}=b_{12}$ represents the impact parameter of the incoming particles, and $n_{12}^i=b_{12}^i/b_{12}$ is the associated unit direction between the particles at that point. Furthermore, we adopt a center-of-mass frame. Since we are already in a small term of order $G$, we can use the special-relativistic notion of center of mass and denote the CM linear momentum by $p^i = m_1 \gamma_1 v_1^i = - m_2 \gamma_2 v_2^i$. In this frame, not only do we have $(b_{12}v_{12})=0$ at the point of minimal approach, but also $(b_{12}v_{1})=(b_{12}v_{2})=0$. The expression~\eqref{DeltaD0} then drastically simplifies, and we obtain
\begin{equation}\label{DeltaD1}
\Delta D^i_{12} = \frac{2}{\sqrt{v_{12}^2+\frac{\gamma_2^2}{c^2}(v_{12}v_2)^2}}\frac{b^i_{12}}{b_{12}^2} \,.
\end{equation}
Note in particular that the scalar products $(u_1\Delta D_{12})$ and $(u_2\Delta D_{12})$ vanish, so only the first term in Eq.~\eqref{changep1} will contribute. On the other hand, we can simplify the result by noting that $v_{12}^2+\frac{\gamma_2^2}{c^2}(v_{12}v_2)^2=c^2\gamma_1^{-2}(\epsilon_{12}^2-1)$. Finally, we obtain the total change in linear momentum during the scattering encounter as
\begin{align}\label{finalchangep1}
\Delta \bm{p}_1 = - \frac{2 G m_1 m_2}{c} \,\frac{2\epsilon_{12}^2-1}{\sqrt{\epsilon_{12}^2-1}}\,\frac{\bm{b}_{12}}{b_{12}^2}\,.
\end{align}
This is a manifestly covariant form [recall that $\epsilon_{12}=-(u_1 u_2)$], which agrees with published results in the literature, like the recent derivation in Fourier space, Eq.~(58) of Ref.~\cite{Dscatt16}. In terms of the center-of-mass linear momentum $\bm{p}=\overline{m}_1 \bm{v}_1=-\overline{m}_2 \bm{v}_2$, where we have posed $\overline{m}_1=\sqrt{m_1^2+\bm{p}^2/c^2}$ and $\overline{m}_2=\sqrt{m_2^2+\bm{p}^2/c^2}$, we have
\begin{align}\label{finalchangep1alt}
\Delta \bm{p}_1 = - \frac{2 G}{\vert\bm{p}\vert} \frac{\overline{m}_1^2 \overline{m}_2^2}{\overline{m}_1 + \overline{m}_2} \left[ 1 + \left( \frac{1}{\overline{m}_1^2} + \frac{1}{\overline{m}_2^2} + \frac{4}{\overline{m}_1 \overline{m}_2}\right)\frac{\bm{p}^2}{c^2} + \frac{\bm{p}^4}{\overline{m}_1^2 \overline{m}_2^2 \,c^4}\right] \frac{\bm{b}_{12}}{b_{12}^2}\,.
\end{align}
The latter form agrees with results from Refs.~\cite{Westpf85,LSB08}. Actually the scattering of two particles has even been computed up to 2PM order in Refs.~\cite{WG79,Port80,WH80,Westpf85}.

\section{Circular orbits and their stability}
\label{app:circ}

Although the most well known physical application of the PM approximation is for unbound orbits, typically in scattering situations (see App.~\ref{app:scatt}), it may be mathematically interesting to investigate the case of bound circular orbits. We suppose that our fundamental theory is not GR but is described by the linearized action~\eqref{S}.

We consider two particles with equal masses $m_1=m_2=\frac{m}{2}$, see the notation in Sec.~\ref{sec:eqmass}. The motion takes place in the fixed orbital plane described by polar coordinates $(r,\varphi)$. Posing $C=r^2 \dot{\varphi}=\vert\bm{x}\times\bm{v}\vert$, the equations of motion, equivalent to those presented in Sec.~\ref{sec:eqmass}, take the form
\begin{subequations}\label{eompolar}
\begin{align}
\ddot{r} - \frac{C^2}{r^3} &= - \frac{G m}{r^2\,y^{3/2}} \left[ 8 \gamma^2 - 8 + \frac{1}{\gamma^2} - \biggl(4\gamma^4 - \frac{1}{2}\biggr) \dot{r}^2 \right] \,, \label{eomradial}\\
\frac{\dot{C}}{C} &= \frac{G m}{r^2\,y^{3/2}} \biggl(4\gamma^4 - \frac{1}{2}\biggr) \dot{r} \,.
\end{align}
\end{subequations}
By integration of these equations we recover the conserved energy $E$ and angular momentum $J$ already obtained in Sec.~\ref{sec:eqmass}, namely
\begin{subequations}\label{EJ}
\begin{align}
E &= m c^2 \gamma + \frac{G m^2}{4 r y^{1/2}} \left(2\gamma^2-1\right)\Bigl(8\gamma^3-8\gamma-\frac{1}{\gamma}\Bigr)\,, \\
J &=  \frac{C}{4} \left[ m\gamma + \frac{G m^2}{r c^2 y^{1/2}} \Bigl(4\gamma^5-2\gamma^3+\frac{3}{2}\gamma-\frac{1}{4\gamma}\Bigr)\right]\,.
\end{align}
\end{subequations}

\begin{figure}[t]
\begin{center}
\begin{tabular}{c}
\includegraphics[width=8.0cm,angle=0]{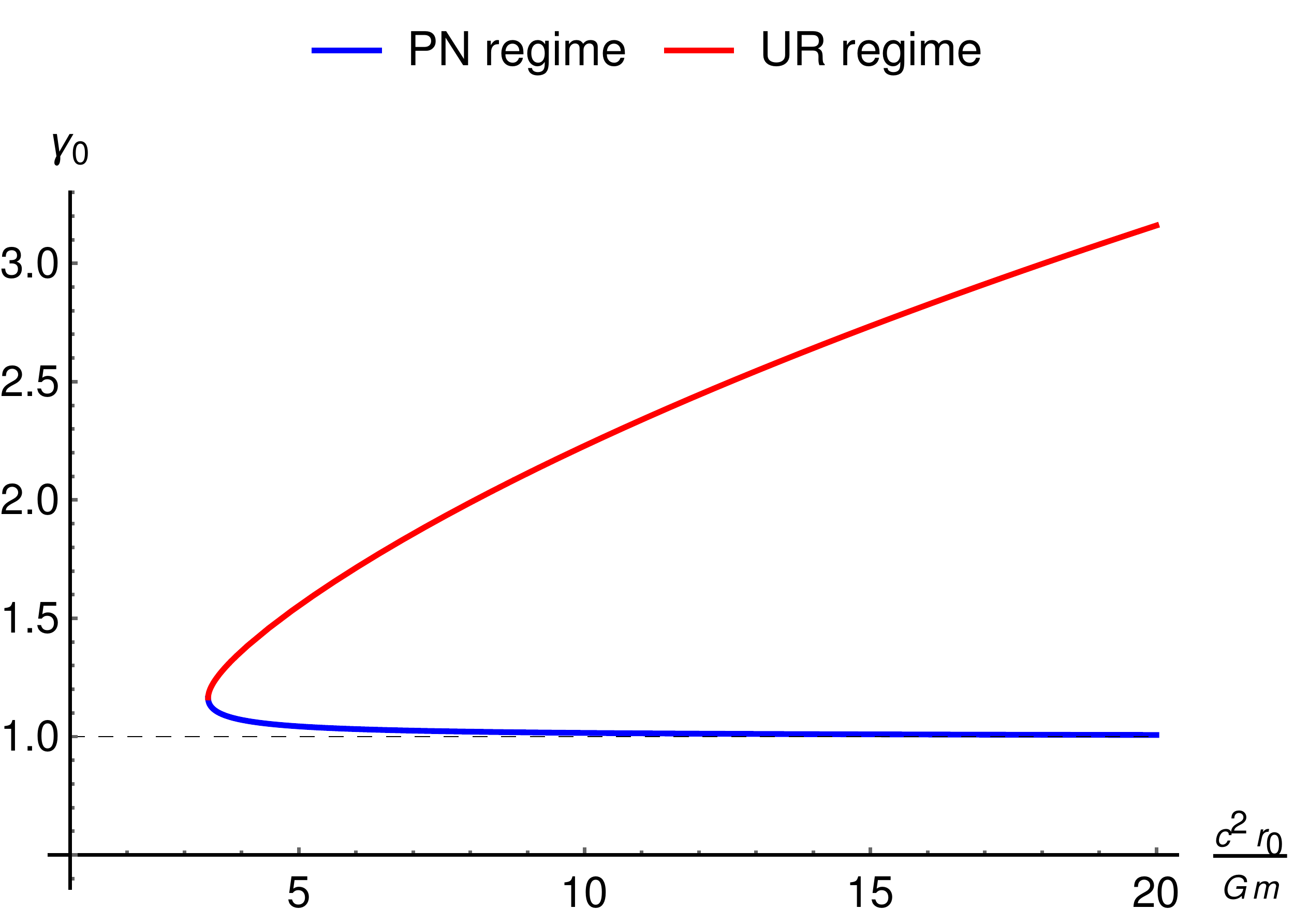}
\hspace{0.8cm}
\includegraphics[width=7.5cm,angle=0]{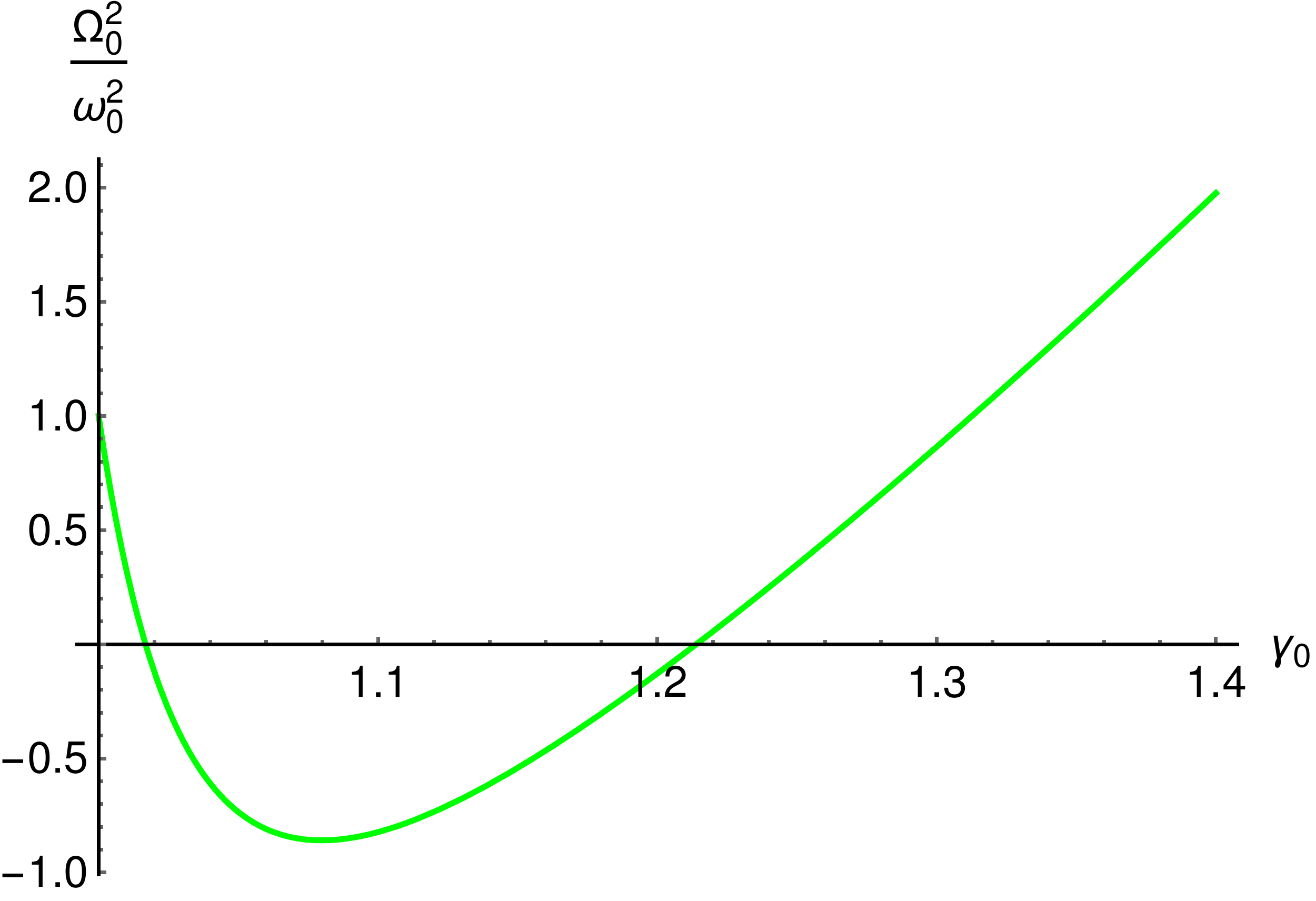}
\end{tabular}
\caption{Left panel: The circular orbits in the two branches corresponding to $\epsilon=1$ (PN regime) and $\epsilon=-1$ (UR regime) in Eq.~\eqref{gamma0}. The point with minimal circular radius corresponds to $r_\text{min} = (2+\sqrt{2})\,\frac{G m}{c^2}$ and $\gamma_\text{min}=\frac{1}{2}(4+\sqrt{2})^{1/2}$. Right panel: Stability criterium of circular orbits. The unstable region ($\Omega_0^2<0$) is for $1.02\lesssim\gamma_0\lesssim 1.21$ and surrounds the minimal radius $r_\text{min}$ in the figure on the left (indeed $\gamma_\text{min}\simeq 1.16$).}\label{fig}
\end{center}
\end{figure}

The circular orbit is defined by $\dot{r}=\ddot{r}=0$, hence $r=r_0$ and $C=C_0$ are constant. We denote the orbital frequency of the circular orbit by $\omega_0=\dot{\varphi}_0=C_0/r_0^2$, given by
\begin{equation}\label{omega0}
\omega_0^2 = \frac{G m}{r_0^3}\Bigl(8 \gamma_0^2 - 8 + \frac{1}{\gamma_0^2}\Bigr)\,.
\end{equation}
Since we have by definition $v_0^2 = r_0^2 \omega_0^2 = 4c^2(1-1/\gamma_0^2)$, the equation~\eqref{omega0} gives the relation linking the orbital frequency $\omega_0$ to the radius $r_0$ for circular orbits. It can also be written as a relation between $u_0=\frac{G m}{r_0 c^2}$ (which represents in fact the small dimensionless PM parameter) and the relativistic factor $\gamma_0$:
\begin{equation}\label{u0}
u_0 = \frac{4(\gamma_0^2-1)}{8 \gamma_0^4 - 8\gamma_0^2 + 1} \,.
\end{equation}
We observe that the limit $u_0\to 0$ (or large separation $r_0\to\infty$) corresponds to two opposite regimes. One is the usual PN limit which is such that $\gamma_0\to 1$. However there is also the ultra-relativistic (UR) regime for which $\gamma_0\to\infty$. In other words the large separation regime is composed of two branches,
\begin{equation}\label{gamma0}
\gamma_0^2 = \frac{1+2u_0-\epsilon \sqrt{1-4u_0+2u_0^2}}{4u_0} \,,
\end{equation}
with $\epsilon=+1$ in the PN regime and $\epsilon=-1$ in the UR regime, see the left panel of Fig.~\ref{fig}. Note that in between these two regimes, there is a minimal value for the radius of a circular orbit, which is given by $r_\text{min} = (2+\sqrt{2})\,\frac{G m}{c^2}$. Along the UR branch $u_0^{-1}$ tends to infinity like $u_0^{-1}=\frac{c^2 r_0}{G m} \sim 2 \gamma_0^2$.

Next we investigate the stability of the circular orbits, under a small linear perturbation [still assuming the linear theory~\eqref{S} which is not GR]. We insert the perturbation ansatz $r=r_0+\delta r$ and $C=C_0+\delta C$ into Eqs.~\eqref{eompolar}, and readily obtain to first order,
\begin{equation}\label{deltaC}
\frac{\delta C}{C_0} = \frac{G m}{r_0^2}\left(4\gamma_0^4-\frac{1}{2}\right)\delta r\,.
\end{equation}
On the other hand, the radial equation arises from a combination of~\eqref{eomradial} with~\eqref{deltaC}, and we find the harmonic oscillator, 
\begin{equation}\label{deltar}
\ddot{\delta r}+\Omega_0^2 \,\delta r = 0\,,
\end{equation}
with characteristic frequency
\begin{equation}\label{Omega0}
\Omega_0^2 = \omega_0^2 \left[ 1 + \frac{2(\gamma_0^2-1)(8\gamma_0^2-1)(16\gamma_0^6-40\gamma_0^4+22\gamma_0^2-1)}{(8\gamma_0^4-8\gamma_0^2+1)^2} \right]\,.
\end{equation}

To study the stability of the circular orbits we just have to look for the sign of the characteristic frequency squared $\Omega_0^2$. From the right panel of Fig.~\ref{fig} we see that $\Omega_0^2$ is negative, and thus the orbit is unstable, when $\gamma_-\leqslant\gamma_0\leqslant\gamma_+$ where $\gamma_- \simeq 1.01678$ and $\gamma_+ \simeq 1.21395$. This corresponds \textit{via} Eq.~\eqref{gamma0} to the point $r_-\sim 9.45\,\frac{G m}{c^2}$ on the PN branch $\epsilon=1$, and to the point $r_+\sim 3.48\,\frac{G m}{c^2}$ on the UR branch $\epsilon=-1$. In particular, we note that the point with minimal circular radius $r_\text{min} \simeq 3.14 \,\frac{G m}{c^2}$ is within the unstable region. 

Therefore, we conclude that the two PN and UR regimes, are in fact separated by the unstable region $\gamma_-\leqslant\gamma_0\leqslant\gamma_+$. This means, for instance, that if we start with a circular orbit in the PN regime with $\gamma_0 \simeq 1$, and increase the velocity we shall never be able to reach the UR regime $\gamma_0 \to \infty$ since the orbit will first become unstable when reaching the point $\gamma_0=\gamma_-$. Similarly, starting in the UR regime it would be impossible to reach the PN regime since the orbit will be unstable when $\gamma_0=\gamma_+$.

\section{The 5PN harmonic-coordinates Lagrangian to order $G$}
\label{app:5PN}

The total time derivative which is to be added to the construction of the 1PM Lagrangian in Sec.~\ref{sec:lagA} in order to ensure the equivalence with the 4PN result~\cite{BBFM17} is of the form
\begin{equation}\label{functionF}
  F_\text{PN} = \frac{1}{c^4}F_\text{2PN} + \frac{1}{c^6}F_\text{3PN} +
  \frac{1}{c^6} F_\text{4PN}\,,
\end{equation}
where the coefficients are explicitly given by
\begin{subequations}\label{FPN}
  \begin{align}
F_\text{2PN} &= \frac{G m_1 m_2}{24} \Bigl[(n_{12}v_1)^2
  (n_{12}v_2)+14 (n_{12}v_1) (v_1v_2)-7 (n_{12}v_2) v_1^2\Bigr] + 1
\leftrightarrow 2\,,\\
F_\text{3PN} &= - \frac{G m_1 m_2}{240} \Bigl[3 (n_{12}v_1)^4
  (n_{12}v_2)+6 (n_{12}v_1)^3 (n_{12}v_2)^2+20 (n_{12}v_1)^3
  (v_1v_2)\nonumber\\&\qquad-15 (n_{12}v_1)^2 (n_{12}v_2) v_1^2+36
  (n_{12}v_1)^2 (n_{12}v_2) (v_1v_2)-24 (n_{12}v_1) (n_{12}v_2)^2
  v_1^2\nonumber\\&\qquad-90 (n_{12}v_1) v_1^2 (v_1v_2)+90 (n_{12}v_1)
  v_1^2 v_2^2+12 (n_{12}v_1) (v_1v_2)^2\nonumber\\&\qquad-30
  (n_{12}v_2)^3 v_1^2-132 (n_{12}v_2) v_1^4+144 (n_{12}v_2) v_1^2
  (v_1v_2)\Bigr] + 1 \leftrightarrow 2\,,\\
F_\text{4PN} &= \frac{G m_1 m_2}{2688} \Bigl[15 (n_{12}v_1)^6
  (n_{12}v_2)+30 (n_{12}v_1)^5 (n_{12}v_2)^2+78 (n_{12}v_1)^5
  (v_1v_2)\nonumber\\&\qquad-60 (n_{12}v_1)^4 (n_{12}v_2)^3-81
  (n_{12}v_1)^4 (n_{12}v_2) v_1^2+132 (n_{12}v_1)^4 (n_{12}v_2)
  (v_1v_2)\nonumber\\&\qquad-132 (n_{12}v_1)^3 (n_{12}v_2)^2 v_1^2+180
  (n_{12}v_1)^3 (n_{12}v_2)^2 (v_1v_2)-308 (n_{12}v_1)^3 v_1^2
  (v_1v_2)\nonumber\\&\qquad+308 (n_{12}v_1)^3 v_1^2 v_2^2-24
  (n_{12}v_1)^3 (v_1v_2)^2-162 (n_{12}v_1)^2 (n_{12}v_2)^3
  v_1^2\nonumber\\&\qquad+159 (n_{12}v_1)^2 (n_{12}v_2) v_1^4-312
  (n_{12}v_1)^2 (n_{12}v_2) v_1^2 (v_1v_2)-552 (n_{12}v_1)^2
  (n_{12}v_2) v_1^2 v_2^2\nonumber\\&\qquad+240 (n_{12}v_1)^2
  (n_{12}v_2) (v_1v_2)^2-180 (n_{12}v_1) (n_{12}v_2)^4 v_1^2+186
  (n_{12}v_1) (n_{12}v_2)^2 v_1^4\nonumber\\&\qquad-252 (n_{12}v_1)
  (n_{12}v_2)^2 v_1^2 (v_1v_2)+738 (n_{12}v_1) v_1^4 (v_1v_2)-738
  (n_{12}v_1) v_1^4 v_2^2\nonumber\\&\qquad-72 (n_{12}v_1) v_1^2
  (v_1v_2)^2+1188 (n_{12}v_1) v_1^2 (v_1v_2) v_2^2-72 (n_{12}v_1)
  (v_1v_2)^3\nonumber\\&\qquad+78 (n_{12}v_2)^5 v_1^2-228
  (n_{12}v_2)^3 v_1^4-272 (n_{12}v_2)^3 v_1^2 (v_1v_2)-225 (n_{12}v_2)
  v_1^6\nonumber\\&\qquad+588 (n_{12}v_2) v_1^4 (v_1v_2)+1116
  (n_{12}v_2) v_1^4 v_2^2+48 (n_{12}v_2) v_1^2 (v_1v_2)^2\Bigr] + 1
\leftrightarrow 2\,.
\end{align}
\end{subequations}

The harmonic-coordinates Lagrangian is specified by the couple $(\lambda, q_a^i)$, where $\lambda$ represents the ordinary part of the Lagrangian and $q_a^i$ denotes the coefficients of the accelerations, see Eq.~\eqref{Ldef}. Up to the 5PN level and linear in $G$, we have
\begin{align}\label{lambdaexp}
\lambda &= \lambda_\text{N} + \frac{1}{c^2}\lambda_\text{1PN} +
\frac{1}{c^4}\lambda_\text{2PN}+ \frac{1}{c^6}\lambda_\text{3PN} +
\frac{1}{c^8}\lambda_\text{4PN}+ \frac{1}{c^{10}}\lambda_\text{5PN} +
\mathcal{O}\left(\frac{1}{c^{12}}\right)\,,\\q_1^i &= \frac{1}{c^4}
q_{1,\text{2PN}}^i + \frac{1}{c^6} q_{1,\text{3PN}}^i + \frac{1}{c^8}
q_{1,\text{4PN}}^i + \frac{1}{c^{10}} q_{1,\text{5PN}}^i +
\mathcal{O}\left(\frac{1}{c^{12}}\right)\,,
\end{align}
together with $q_2^i$ obtained by label exchange. We explicitly find, after adjustement with the function $F_\text{PN}$ obtained previously,
\begin{subequations}\label{lambdaPN}
  \begin{align}
    \lambda_\text{N} &= \frac{G m_1 m_2}{r_{12}} \,,\\
\lambda_\text{1PN} &= - \frac{G m_1 m_2}{4 r_{12}}\Bigl[(n_{12}v_1)
  (n_{12}v_2)-6 v_1^2+7 (v_1v_2)\Bigr] + 1 \leftrightarrow 2\,,\\
\lambda_\text{2PN} &= \frac{G m_1 m_2}{16 r_{12}}\Bigl[3 (n_{12}v_1)^2
  (n_{12}v_2)^2+12 (n_{12}v_1) (n_{12}v_2) (v_1v_2)-14 (n_{12}v_2)^2
  v_1^2\nonumber\\&\qquad+14 v_1^4-32 v_1^2 (v_1v_2)+15 v_1^2 v_2^2+2
  (v_1v_2)^2\Bigr] + 1 \leftrightarrow 2\,,\\
\lambda_\text{3PN} &= - \frac{G m_1 m_2}{32 r_{12}}\Bigl[5
  (n_{12}v_1)^3 (n_{12}v_2)^3+15 (n_{12}v_1)^2 (n_{12}v_2)^2
  (v_1v_2)+20 (n_{12}v_1)^2 v_1^2 v_2^2\nonumber\\&\qquad-4
  (n_{12}v_1) (n_{12}v_2)^3 v_1^2+22 (n_{12}v_1) (n_{12}v_2) v_1^4-32
  (n_{12}v_1) (n_{12}v_2) v_1^2 (v_1v_2)\nonumber\\&\qquad+23
  (n_{12}v_1) (n_{12}v_2) v_1^2 v_2^2-10 (n_{12}v_1) (n_{12}v_2)
  (v_1v_2)^2-20 (n_{12}v_2)^4 v_1^2\nonumber\\&\qquad-8 (n_{12}v_2)^2
  v_1^4-12 (n_{12}v_2)^2 v_1^2 (v_1v_2)-22 v_1^6+26 v_1^4 (v_1v_2)-2
  v_1^4 v_2^2\nonumber\\&\qquad+v_1^2 (v_1v_2) v_2^2-2
  (v_1v_2)^3\Bigr] + 1 \leftrightarrow 2\,,\\
\lambda_\text{4PN} &= \frac{G m_1 m_2}{256 r_{12}}\Bigl[70
  (n_{12}v_1)^5 (n_{12}v_2)^3-35 (n_{12}v_1)^4 (n_{12}v_2)^4-30
  (n_{12}v_1)^4 (n_{12}v_2)^2 (v_1v_2)\nonumber\\&\qquad-40
  (n_{12}v_1)^3 (n_{12}v_2)^3 v_1^2+120 (n_{12}v_1)^3 (n_{12}v_2)^3
  (v_1v_2)+396 (n_{12}v_1)^3 (n_{12}v_2) v_1^2
  v_2^2\nonumber\\&\qquad-120 (n_{12}v_1)^3 (n_{12}v_2) (v_1v_2)^2-84
  (n_{12}v_1)^2 (n_{12}v_2)^2 v_1^2 v_2^2+44 (n_{12}v_1)^2 v_1^2
  (v_1v_2) v_2^2\nonumber\\&\qquad+40 (n_{12}v_1)^2 (v_1v_2)^3-150
  (n_{12}v_1) (n_{12}v_2)^5 v_1^2+114 (n_{12}v_1) (n_{12}v_2)^3
  v_1^4\nonumber\\&\qquad-80 (n_{12}v_1) (n_{12}v_2)^3 v_1^2
  (v_1v_2)-370 (n_{12}v_1) (n_{12}v_2) v_1^4 v_2^2+72 (n_{12}v_1)
  (n_{12}v_2) v_1^2 (v_1v_2)^2\nonumber\\&\qquad+280 (n_{12}v_1)
  (n_{12}v_2) v_1^2 (v_1v_2) v_2^2-48 (n_{12}v_1) (n_{12}v_2)
  (v_1v_2)^3-30 (n_{12}v_2)^4 v_1^4\nonumber\\&\qquad-106
  (n_{12}v_2)^4 v_1^2 (v_1v_2)-88 (n_{12}v_2)^2 v_1^6+78 (n_{12}v_2)^2
  v_1^4 (v_1v_2)+92 (n_{12}v_2)^2 v_1^4 v_2^2\nonumber\\&\qquad-120
  (n_{12}v_2)^2 v_1^2 (v_1v_2)^2+150 v_1^8-320 v_1^6 (v_1v_2)+160
  v_1^6 v_2^2+192 v_1^4 (v_1v_2)^2\nonumber\\&\qquad-198 v_1^4
  (v_1v_2) v_2^2+3 v_1^4 v_2^4+8 v_1^2 (v_1v_2)^3-8 v_1^2 (v_1v_2)^2
  v_2^2+8 (v_1v_2)^4\Bigr] + 1 \leftrightarrow 2\,,\\
\lambda_\text{5PN} &= - \frac{G m_1 m_2}{4608 r_{12}}\Bigl[126
  (n_{12}v_2) (n_{12}v_1)^9+126 (n_{12}v_2)^2 (n_{12}v_1)^8+434
  (v_1v_2) (n_{12}v_1)^8\nonumber\\&\qquad+126 (n_{12}v_2)^3
  (n_{12}v_1)^7-714 (n_{12}v_2) v_1^2 (n_{12}v_1)^7+336 (n_{12}v_2)
  (v_1v_2) (n_{12}v_1)^7\nonumber\\&\qquad+126 (n_{12}v_2)^4
  (n_{12}v_1)^6+192 (v_1v_2)^2 (n_{12}v_1)^6-602 (n_{12}v_2)^2 v_1^2
  (n_{12}v_1)^6\nonumber\\&\qquad+266 (n_{12}v_2)^2 (v_1v_2)
  (n_{12}v_1)^6-1846 v_1^2 (v_1v_2) (n_{12}v_1)^6+1846 v_1^2 v_2^2
  (n_{12}v_1)^6\nonumber\\&\qquad+63 (n_{12}v_2)^5 (n_{12}v_1)^5+1526
  (n_{12}v_2) v_1^4 (n_{12}v_1)^5-388 (n_{12}v_2) (v_1v_2)^2
  (n_{12}v_1)^5\nonumber\\&\qquad-504 (n_{12}v_2)^3 v_1^2
  (n_{12}v_1)^5+224 (n_{12}v_2)^3 (v_1v_2) (n_{12}v_1)^5-968
  (n_{12}v_2) v_1^2 (v_1v_2) (n_{12}v_1)^5\nonumber\\&\qquad+886
  (n_{12}v_2) v_1^2 v_2^2 (n_{12}v_1)^5+76 (v_1v_2)^3
  (n_{12}v_1)^4+1010 (n_{12}v_2)^2 v_1^4
  (n_{12}v_1)^4\nonumber\\&\qquad-520 (n_{12}v_2)^2 (v_1v_2)^2
  (n_{12}v_1)^4-820 v_1^2 (v_1v_2)^2 (n_{12}v_1)^4-420 (n_{12}v_2)^4
  v_1^2 (n_{12}v_1)^4\nonumber\\&\qquad+105 (n_{12}v_2)^4 (v_1v_2)
  (n_{12}v_1)^4+3082 v_1^4 (v_1v_2) (n_{12}v_1)^4-440 (n_{12}v_2)^2
  v_1^2 (v_1v_2) (n_{12}v_1)^4\nonumber\\&\qquad-3082 v_1^4 v_2^2
  (n_{12}v_1)^4+850 (n_{12}v_2)^2 v_1^2 v_2^2 (n_{12}v_1)^4+2874 v_1^2
  (v_1v_2) v_2^2 (n_{12}v_1)^4\nonumber\\&\qquad-1546 (n_{12}v_2)
  v_1^6 (n_{12}v_1)^3+256 (n_{12}v_2) (v_1v_2)^3 (n_{12}v_1)^3+650
  (n_{12}v_2)^3 v_1^4 (n_{12}v_1)^3\nonumber\\&\qquad-280
  (n_{12}v_2)^3 (v_1v_2)^2 (n_{12}v_1)^3+736 (n_{12}v_2) v_1^2
  (v_1v_2)^2 (n_{12}v_1)^3-350 (n_{12}v_2)^5 v_1^2
  (n_{12}v_1)^3\nonumber\\&\qquad+992 (n_{12}v_2) v_1^4 (v_1v_2)
  (n_{12}v_1)^3-160 (n_{12}v_2)^3 v_1^2 (v_1v_2) (n_{12}v_1)^3-1160
  (n_{12}v_2) v_1^4 v_2^2 (n_{12}v_1)^3\nonumber\\&\qquad+420
  (n_{12}v_2)^3 v_1^2 v_2^2 (n_{12}v_1)^3+144 (n_{12}v_2) v_1^2
  (v_1v_2) v_2^2 (n_{12}v_1)^3-16 (v_1v_2)^4
  (n_{12}v_1)^2\nonumber\\&\qquad-738 (n_{12}v_2)^2 v_1^6
  (n_{12}v_1)^2+168 (n_{12}v_2)^2 (v_1v_2)^3 (n_{12}v_1)^2-96 v_1^2
  (v_1v_2)^3 (n_{12}v_1)^2\nonumber\\&\qquad+410 (n_{12}v_2)^4 v_1^4
  (n_{12}v_1)^2+1272 v_1^4 (v_1v_2)^2 (n_{12}v_1)^2+576 (n_{12}v_2)^2
  v_1^2 (v_1v_2)^2 (n_{12}v_1)^2\nonumber\\&\qquad+2640 v_1^4 v_2^4
  (n_{12}v_1)^2-294 (n_{12}v_2)^6 v_1^2 (n_{12}v_1)^2-2550 v_1^6
  (v_1v_2) (n_{12}v_1)^2\nonumber\\&\qquad+222 (n_{12}v_2)^2 v_1^4
  (v_1v_2) (n_{12}v_1)^2-50 (n_{12}v_2)^4 v_1^2 (v_1v_2)
  (n_{12}v_1)^2+2550 v_1^6 v_2^2 (n_{12}v_1)^2\nonumber\\&\qquad-870
  (n_{12}v_2)^2 v_1^4 v_2^2 (n_{12}v_1)^2+1416 v_1^2 (v_1v_2)^2 v_2^2
  (n_{12}v_1)^2-3912 v_1^4 (v_1v_2) v_2^2
  (n_{12}v_1)^2\nonumber\\&\qquad+12 (n_{12}v_2)^2 v_1^2 (v_1v_2)
  v_2^2 (n_{12}v_1)^2+758 (n_{12}v_2) v_1^8 (n_{12}v_1)-24 (n_{12}v_2)
  (v_1v_2)^4 (n_{12}v_1)\nonumber\\&\qquad-348 (n_{12}v_2)^3 v_1^6
  (n_{12}v_1)-192 (n_{12}v_2) v_1^2 (v_1v_2)^3 (n_{12}v_1)+260
  (n_{12}v_2)^5 v_1^4 (n_{12}v_1)\nonumber\\&\qquad-372 (n_{12}v_2)
  v_1^4 (v_1v_2)^2 (n_{12}v_1)+296 (n_{12}v_2)^3 v_1^2 (v_1v_2)^2
  (n_{12}v_1)+351 (n_{12}v_2) v_1^4 v_2^4
  (n_{12}v_1)\nonumber\\&\qquad-252 (n_{12}v_2)^7 v_1^2
  (n_{12}v_1)-504 (n_{12}v_2) v_1^6 (v_1v_2) (n_{12}v_1)+64
  (n_{12}v_2)^3 v_1^4 (v_1v_2) (n_{12}v_1)\nonumber\\&\qquad-56
  (n_{12}v_2)^5 v_1^2 (v_1v_2) (n_{12}v_1)+730 (n_{12}v_2) v_1^6 v_2^2
  (n_{12}v_1)-694 (n_{12}v_2)^3 v_1^4 v_2^2
  (n_{12}v_1)\nonumber\\&\qquad-168 (n_{12}v_2) v_1^2 (v_1v_2)^2 v_2^2
  (n_{12}v_1)-264 (n_{12}v_2) v_1^4 (v_1v_2) v_2^2 (n_{12}v_1)-2394
  v_1^{10}\nonumber\\&\qquad-8 (v_1v_2)^5+1316 (n_{12}v_2)^2 v_1^8-32
  v_1^2 (v_1v_2)^4-844 (n_{12}v_2)^4 v_1^6-36 v_1^4
  (v_1v_2)^3\nonumber\\&\qquad-56 (n_{12}v_2)^2 v_1^2 (v_1v_2)^3+596
  (n_{12}v_2)^6 v_1^4-3292 v_1^6 (v_1v_2)^2+888 (n_{12}v_2)^2 v_1^4
  (v_1v_2)^2\nonumber\\&\qquad-524 (n_{12}v_2)^4 v_1^2 (v_1v_2)^2-3450
  v_1^6 v_2^4+3465 v_1^4 (v_1v_2) v_2^4-434 (n_{12}v_2)^8
  v_1^2\nonumber\\&\qquad+5770 v_1^8 (v_1v_2)-2132 (n_{12}v_2)^2 v_1^6
  (v_1v_2)+1292 (n_{12}v_2)^4 v_1^4 (v_1v_2)-788 (n_{12}v_2)^6 v_1^2
  (v_1v_2)\nonumber\\&\qquad-3250 v_1^8 v_2^2+2178 (n_{12}v_2)^2 v_1^6
  v_2^2-8 v_1^2 (v_1v_2)^3 v_2^2-2054 (n_{12}v_2)^4 v_1^4
  v_2^2\nonumber\\&\qquad-3444 v_1^4 (v_1v_2)^2 v_2^2+6742 v_1^6
  (v_1v_2) v_2^2-3498 (n_{12}v_2)^2 v_1^4 (v_1v_2) v_2^2\Bigr] + 1
\leftrightarrow 2\,,
\end{align}
\end{subequations}
and
\begin{subequations}\label{q1PN}
  \begin{align}
q_{1,\text{2PN}}^i &= - \frac{G m_1 m_2}{8}\Bigl[\Bigl((n_{12}v_2)^2-7
  v_2^2\Bigr)n_{12}^i+14 (n_{12}v_2) v_{2}^i\Bigr] \,,\\
q_{1,\text{3PN}}^i &= - \frac{G m_1 m_2}{48}\Bigl[\Bigl(30
  (n_{12}v_1)^2 v_2^2-6 (n_{12}v_1) (n_{12}v_2)^3+24 (n_{12}v_1)
  (n_{12}v_2) v_2^2\nonumber\\&\qquad-3 (n_{12}v_2)^4-18 (n_{12}v_2)^2
  (v_1v_2)+15 (n_{12}v_2)^2 v_2^2\Bigr)n_{12}^i \nonumber\\&\quad
  +\Bigl(-132 (n_{12}v_2) v_1^2+96 (n_{12}v_2) (v_1v_2)-90 (n_{12}v_2)
  v_2^2\Bigr)v_{1}^i\nonumber\\&\quad +\Bigl(-18 (n_{12}v_1)
  (n_{12}v_2)^2-20 (n_{12}v_2)^3+48 (n_{12}v_2) v_1^2-12 (n_{12}v_2)
  (v_1v_2)+90 (n_{12}v_2) v_2^2\Bigr)v_{2}^i\Bigr] \,,\\
q_{1,\text{4PN}}^i &= - \frac{G m_1 m_2}{384}\Bigl[\Bigl(60
  (n_{12}v_1)^3 (n_{12}v_2)^3-180 (n_{12}v_1)^3 (n_{12}v_2) v_2^2-162
  (n_{12}v_1)^2 (n_{12}v_2)^2 v_2^2\nonumber\\&\qquad-204
  (n_{12}v_1)^2 (v_1v_2) v_2^2+30 (n_{12}v_1) (n_{12}v_2)^5+120
  (n_{12}v_1) (n_{12}v_2)^3 (v_1v_2)\nonumber\\&\qquad-132 (n_{12}v_1)
  (n_{12}v_2)^3 v_2^2+276 (n_{12}v_1) (n_{12}v_2) v_1^2 v_2^2-120
  (n_{12}v_1) (n_{12}v_2) (v_1v_2)^2\nonumber\\&\qquad-168 (n_{12}v_1)
  (n_{12}v_2) (v_1v_2) v_2^2+186 (n_{12}v_1) (n_{12}v_2) v_2^4+15
  (n_{12}v_2)^6\nonumber\\&\qquad+66 (n_{12}v_2)^4 (v_1v_2)-81
  (n_{12}v_2)^4 v_2^2-156 (n_{12}v_2)^2 (v_1v_2)
  v_2^2\nonumber\\&\qquad+159 (n_{12}v_2)^2 v_2^4+294 (v_1v_2)
  v_2^4-225 v_2^6\Bigr)n_{12}^i \nonumber\\&\quad +\Bigl(276
  (n_{12}v_1)^2 (n_{12}v_2) v_2^2-78 (n_{12}v_2)^5+228 (n_{12}v_2)^3
  v_1^2\nonumber\\&\qquad+308 (n_{12}v_2)^3 v_2^2-1116 (n_{12}v_2)
  v_1^2 v_2^2-24 (n_{12}v_2) (v_1v_2)^2\nonumber\\&\qquad+792
  (n_{12}v_2) (v_1v_2) v_2^2-738 (n_{12}v_2)
  v_2^4\Bigr)v_{1}^i\nonumber\\&\quad +\Bigl(-68 (n_{12}v_1)^3
  v_2^2+60 (n_{12}v_1)^2 (n_{12}v_2)^3-120 (n_{12}v_1)^2 (n_{12}v_2)
  (v_1v_2)\nonumber\\&\qquad-84 (n_{12}v_1)^2 (n_{12}v_2) v_2^2+66
  (n_{12}v_1) (n_{12}v_2)^4-156 (n_{12}v_1) (n_{12}v_2)^2
  v_2^2\nonumber\\&\qquad+294 (n_{12}v_1) v_2^4+78 (n_{12}v_2)^5-24
  (n_{12}v_2)^3 (v_1v_2)\nonumber\\&\qquad-308 (n_{12}v_2)^3 v_2^2-24
  (n_{12}v_2) v_1^2 (v_1v_2)+396 (n_{12}v_2) v_1^2
  v_2^2\nonumber\\&\qquad-72 (n_{12}v_2) (v_1v_2)^2-72 (n_{12}v_2)
  (v_1v_2) v_2^2+738 (n_{12}v_2) v_2^4\Bigr)v_{2}^i\Bigr]\,,\\
q_{1,\text{5PN}}^i &= \frac{G m_1 m_2}{5760}\Bigl[\Bigl(140
  (n_{12}v_1)^7 (n_{12}v_2)+245 (n_{12}v_1)^6 (n_{12}v_2)^2+560
  (n_{12}v_1)^6 (v_1v_2)\nonumber\\&\qquad-560 (n_{12}v_1)^6 v_2^2+315
  (n_{12}v_1)^5 (n_{12}v_2)^3-645 (n_{12}v_1)^5 (n_{12}v_2)
  v_1^2\nonumber\\&\qquad+780 (n_{12}v_1)^5 (n_{12}v_2) (v_1v_2)-810
  (n_{12}v_1)^5 (n_{12}v_2) v_2^2+350 (n_{12}v_1)^4
  (n_{12}v_2)^4\nonumber\\&\qquad-900 (n_{12}v_1)^4 (n_{12}v_2)^2
  v_1^2+825 (n_{12}v_1)^4 (n_{12}v_2)^2 (v_1v_2)-1050 (n_{12}v_1)^4
  (n_{12}v_2)^2 v_2^2\nonumber\\&\qquad-1855 (n_{12}v_1)^4 v_1^2
  (v_1v_2)+1855 (n_{12}v_1)^4 v_1^2 v_2^2-190 (n_{12}v_1)^4
  (v_1v_2)^2\nonumber\\&\qquad-690 (n_{12}v_1)^4 (v_1v_2) v_2^2+880
  (n_{12}v_1)^4 v_2^4+350 (n_{12}v_1)^3
  (n_{12}v_2)^5\nonumber\\&\qquad-900 (n_{12}v_1)^3 (n_{12}v_2)^3
  v_1^2+800 (n_{12}v_1)^3 (n_{12}v_2)^3 (v_1v_2)-1250 (n_{12}v_1)^3
  (n_{12}v_2)^3 v_2^2\nonumber\\&\qquad+1010 (n_{12}v_1)^3 (n_{12}v_2)
  v_1^4-1600 (n_{12}v_1)^3 (n_{12}v_2) v_1^2 (v_1v_2)+2010
  (n_{12}v_1)^3 (n_{12}v_2) v_1^2 v_2^2\nonumber\\&\qquad-900
  (n_{12}v_1)^3 (n_{12}v_2) (v_1v_2)^2-560 (n_{12}v_1)^3 (n_{12}v_2)
  (v_1v_2) v_2^2+1300 (n_{12}v_1)^3 (n_{12}v_2)
  v_2^4\nonumber\\&\qquad+315 (n_{12}v_1)^2 (n_{12}v_2)^6-750
  (n_{12}v_1)^2 (n_{12}v_2)^4 v_1^2+750 (n_{12}v_1)^2 (n_{12}v_2)^4
  (v_1v_2)\nonumber\\&\qquad-1350 (n_{12}v_1)^2 (n_{12}v_2)^4
  v_2^2+975 (n_{12}v_1)^2 (n_{12}v_2)^2 v_1^4-900 (n_{12}v_1)^2
  (n_{12}v_2)^2 v_1^2 (v_1v_2)\nonumber\\&\qquad+1920 (n_{12}v_1)^2
  (n_{12}v_2)^2 v_1^2 v_2^2-1200 (n_{12}v_1)^2 (n_{12}v_2)^2
  (v_1v_2)^2-750 (n_{12}v_1)^2 (n_{12}v_2)^2 (v_1v_2)
  v_2^2\nonumber\\&\qquad+1845 (n_{12}v_1)^2 (n_{12}v_2)^2 v_2^4+2250
  (n_{12}v_1)^2 v_1^4 (v_1v_2)-2250 (n_{12}v_1)^2 v_1^4
  v_2^2\nonumber\\&\qquad+240 (n_{12}v_1)^2 v_1^2 (v_1v_2)^2+1950
  (n_{12}v_1)^2 v_1^2 (v_1v_2) v_2^2-2190 (n_{12}v_1)^2 v_1^2
  v_2^4\nonumber\\&\qquad+60 (n_{12}v_1)^2 (v_1v_2)^3+120
  (n_{12}v_1)^2 (v_1v_2)^2 v_2^2+1200 (n_{12}v_1)^2 (v_1v_2)
  v_2^4\nonumber\\&\qquad-1380 (n_{12}v_1)^2 v_2^6+245 (n_{12}v_1)
  (n_{12}v_2)^7-525 (n_{12}v_1) (n_{12}v_2)^5
  v_1^2\nonumber\\&\qquad+660 (n_{12}v_1) (n_{12}v_2)^5 (v_1v_2)-1260
  (n_{12}v_1) (n_{12}v_2)^5 v_2^2+615 (n_{12}v_1) (n_{12}v_2)^3
  v_1^4\nonumber\\&\qquad-400 (n_{12}v_1) (n_{12}v_2)^3 v_1^2
  (v_1v_2)+1600 (n_{12}v_1) (n_{12}v_2)^3 v_1^2 v_2^2-1000 (n_{12}v_1)
  (n_{12}v_2)^3 (v_1v_2)^2\nonumber\\&\qquad-1200 (n_{12}v_1)
  (n_{12}v_2)^3 (v_1v_2) v_2^2+2275 (n_{12}v_1) (n_{12}v_2)^3
  v_2^4-615 (n_{12}v_1) (n_{12}v_2) v_1^6\nonumber\\&\qquad+1020
  (n_{12}v_1) (n_{12}v_2) v_1^4 (v_1v_2)-1560 (n_{12}v_1) (n_{12}v_2)
  v_1^4 v_2^2+840 (n_{12}v_1) (n_{12}v_2) v_1^2
  (v_1v_2)^2\nonumber\\&\qquad+960 (n_{12}v_1) (n_{12}v_2) v_1^2
  (v_1v_2) v_2^2-2025 (n_{12}v_1) (n_{12}v_2) v_1^2 v_2^4+320
  (n_{12}v_1) (n_{12}v_2) (v_1v_2)^3\nonumber\\&\qquad+900 (n_{12}v_1)
  (n_{12}v_2) (v_1v_2)^2 v_2^2+1140 (n_{12}v_1) (n_{12}v_2) (v_1v_2)
  v_2^4-2030 (n_{12}v_1) (n_{12}v_2) v_2^6\nonumber\\&\qquad+140
  (n_{12}v_2)^8-270 (n_{12}v_2)^6 v_1^2+455 (n_{12}v_2)^6
  (v_1v_2)\nonumber\\&\qquad-860 (n_{12}v_2)^6 v_2^2+260 (n_{12}v_2)^4
  v_1^4-175 (n_{12}v_2)^4 v_1^2 (v_1v_2)\nonumber\\&\qquad+1005
  (n_{12}v_2)^4 v_1^2 v_2^2-450 (n_{12}v_2)^4 (v_1v_2)^2-1400
  (n_{12}v_2)^4 (v_1v_2) v_2^2\nonumber\\&\qquad+2020 (n_{12}v_2)^4
  v_2^4-290 (n_{12}v_2)^2 v_1^6+285 (n_{12}v_2)^2 v_1^4
  (v_1v_2)\nonumber\\&\qquad-810 (n_{12}v_2)^2 v_1^4 v_2^2+360
  (n_{12}v_2)^2 v_1^2 (v_1v_2)^2+600 (n_{12}v_2)^2 v_1^2 (v_1v_2)
  v_2^2\nonumber\\&\qquad-1560 (n_{12}v_2)^2 v_1^2 v_2^4+200
  (n_{12}v_2)^2 (v_1v_2)^3+840 (n_{12}v_2)^2 (v_1v_2)^2
  v_2^2\nonumber\\&\qquad+1785 (n_{12}v_2)^2 (v_1v_2) v_2^4-2460
  (n_{12}v_2)^2 v_2^6-1245 v_1^6 (v_1v_2)+1245 v_1^6
  v_2^2\nonumber\\&\qquad+90 v_1^4 (v_1v_2)^2-1980 v_1^4 (v_1v_2)
  v_2^2+1890 v_1^4 v_2^4+120 v_1^2 (v_1v_2)^3-2475 v_1^2 (v_1v_2)
  v_2^4\nonumber\\&\qquad+2355 v_1^2 v_2^6+80 (v_1v_2)^4+100
  (v_1v_2)^3 v_2^2+30 (v_1v_2)^2 v_2^4-2870 (v_1v_2) v_2^6+2660
  v_2^8\Bigr)n_{12}^i \nonumber\\&\quad +\Bigl(-215 (n_{12}v_1)^6
  (n_{12}v_2)-360 (n_{12}v_1)^5 (n_{12}v_2)^2-742 (n_{12}v_1)^5
  (v_1v_2)\nonumber\\&\qquad+742 (n_{12}v_1)^5 v_2^2-450 (n_{12}v_1)^4
  (n_{12}v_2)^3+1010 (n_{12}v_1)^4 (n_{12}v_2)
  v_1^2\nonumber\\&\qquad-800 (n_{12}v_1)^4 (n_{12}v_2) (v_1v_2)+1005
  (n_{12}v_1)^4 (n_{12}v_2) v_2^2-500 (n_{12}v_1)^3
  (n_{12}v_2)^4\nonumber\\&\qquad+1300 (n_{12}v_1)^3 (n_{12}v_2)^2
  v_1^2-600 (n_{12}v_1)^3 (n_{12}v_2)^2 (v_1v_2)+1280 (n_{12}v_1)^3
  (n_{12}v_2)^2 v_2^2\nonumber\\&\qquad+3000 (n_{12}v_1)^3 v_1^2
  (v_1v_2)-3000 (n_{12}v_1)^3 v_1^2 v_2^2+160 (n_{12}v_1)^3
  (v_1v_2)^2\nonumber\\&\qquad+1300 (n_{12}v_1)^3 (v_1v_2) v_2^2-1460
  (n_{12}v_1)^3 v_2^4-525 (n_{12}v_1)^2
  (n_{12}v_2)^5\nonumber\\&\qquad+1230 (n_{12}v_1)^2 (n_{12}v_2)^3
  v_1^2-400 (n_{12}v_1)^2 (n_{12}v_2)^3 (v_1v_2)+1600 (n_{12}v_1)^2
  (n_{12}v_2)^3 v_2^2\nonumber\\&\qquad-1845 (n_{12}v_1)^2 (n_{12}v_2)
  v_1^4+2040 (n_{12}v_1)^2 (n_{12}v_2) v_1^2 (v_1v_2)-3120
  (n_{12}v_1)^2 (n_{12}v_2) v_1^2 v_2^2\nonumber\\&\qquad+840
  (n_{12}v_1)^2 (n_{12}v_2) (v_1v_2)^2+960 (n_{12}v_1)^2 (n_{12}v_2)
  (v_1v_2) v_2^2-2025 (n_{12}v_1)^2 (n_{12}v_2)
  v_2^4\nonumber\\&\qquad-540 (n_{12}v_1) (n_{12}v_2)^6+1040
  (n_{12}v_1) (n_{12}v_2)^4 v_1^2-350 (n_{12}v_1) (n_{12}v_2)^4
  (v_1v_2)\nonumber\\&\qquad+2010 (n_{12}v_1) (n_{12}v_2)^4 v_2^2-1740
  (n_{12}v_1) (n_{12}v_2)^2 v_1^4+1140 (n_{12}v_1) (n_{12}v_2)^2 v_1^2
  (v_1v_2)\nonumber\\&\qquad-3240 (n_{12}v_1) (n_{12}v_2)^2 v_1^2
  v_2^2+720 (n_{12}v_1) (n_{12}v_2)^2 (v_1v_2)^2+1200 (n_{12}v_1)
  (n_{12}v_2)^2 (v_1v_2) v_2^2\nonumber\\&\qquad-3120 (n_{12}v_1)
  (n_{12}v_2)^2 v_2^4-7470 (n_{12}v_1) v_1^4 (v_1v_2)+7470 (n_{12}v_1)
  v_1^4 v_2^2\nonumber\\&\qquad+360 (n_{12}v_1) v_1^2 (v_1v_2)^2-7920
  (n_{12}v_1) v_1^2 (v_1v_2) v_2^2+7560 (n_{12}v_1) v_1^2
  v_2^4\nonumber\\&\qquad+240 (n_{12}v_1) (v_1v_2)^3-4950 (n_{12}v_1)
  (v_1v_2) v_2^4+4710 (n_{12}v_1) v_2^6\nonumber\\&\qquad-560
  (n_{12}v_2)^7+880 (n_{12}v_2)^5 v_1^2-552 (n_{12}v_2)^5
  (v_1v_2)\nonumber\\&\qquad+2597 (n_{12}v_2)^5 v_2^2-1380
  (n_{12}v_2)^3 v_1^4+1280 (n_{12}v_2)^3 v_1^2
  (v_1v_2)\nonumber\\&\qquad-3650 (n_{12}v_2)^3 v_1^2 v_2^2+100
  (n_{12}v_2)^3 (v_1v_2)^2+2600 (n_{12}v_2)^3 (v_1v_2)
  v_2^2\nonumber\\&\qquad-5250 (n_{12}v_2)^3 v_2^4+2660 (n_{12}v_2)
  v_1^6-4920 (n_{12}v_2) v_1^4 (v_1v_2)\nonumber\\&\qquad+7065
  (n_{12}v_2) v_1^4 v_2^2+60 (n_{12}v_2) v_1^2 (v_1v_2)^2-7920
  (n_{12}v_2) v_1^2 (v_1v_2) v_2^2\nonumber\\&\qquad+9450 (n_{12}v_2)
  v_1^2 v_2^4+160 (n_{12}v_2) (v_1v_2)^3-7920 (n_{12}v_2) (v_1v_2)
  v_2^4+8715 (n_{12}v_2) v_2^6\Bigr)v_{1}^i\nonumber\\&\quad +\Bigl(80
  (n_{12}v_1)^7+130 (n_{12}v_1)^6 (n_{12}v_2)+165 (n_{12}v_1)^5
  (n_{12}v_2)^2-371 (n_{12}v_1)^5 v_1^2\nonumber\\&\qquad-76
  (n_{12}v_1)^5 (v_1v_2)-138 (n_{12}v_1)^5 v_2^2+200 (n_{12}v_1)^4
  (n_{12}v_2)^3\nonumber\\&\qquad-400 (n_{12}v_1)^4 (n_{12}v_2)
  v_1^2-450 (n_{12}v_1)^4 (n_{12}v_2) (v_1v_2)-140 (n_{12}v_1)^4
  (n_{12}v_2) v_2^2\nonumber\\&\qquad+250 (n_{12}v_1)^3
  (n_{12}v_2)^4-300 (n_{12}v_1)^3 (n_{12}v_2)^2 v_1^2-800
  (n_{12}v_1)^3 (n_{12}v_2)^2 (v_1v_2)\nonumber\\&\qquad-250
  (n_{12}v_1)^3 (n_{12}v_2)^2 v_2^2+750 (n_{12}v_1)^3 v_1^4+160
  (n_{12}v_1)^3 v_1^2 (v_1v_2)\nonumber\\&\qquad+650 (n_{12}v_1)^3
  v_1^2 v_2^2+60 (n_{12}v_1)^3 (v_1v_2)^2+80 (n_{12}v_1)^3 (v_1v_2)
  v_2^2\nonumber\\&\qquad+400 (n_{12}v_1)^3 v_2^4+330 (n_{12}v_1)^2
  (n_{12}v_2)^5-200 (n_{12}v_1)^2 (n_{12}v_2)^3
  v_1^2\nonumber\\&\qquad-1000 (n_{12}v_1)^2 (n_{12}v_2)^3
  (v_1v_2)-600 (n_{12}v_1)^2 (n_{12}v_2)^3 v_2^2+510 (n_{12}v_1)^2
  (n_{12}v_2) v_1^4\nonumber\\&\qquad+840 (n_{12}v_1)^2 (n_{12}v_2)
  v_1^2 (v_1v_2)+480 (n_{12}v_1)^2 (n_{12}v_2) v_1^2 v_2^2+480
  (n_{12}v_1)^2 (n_{12}v_2) (v_1v_2)^2\nonumber\\&\qquad+900
  (n_{12}v_1)^2 (n_{12}v_2) (v_1v_2) v_2^2+570 (n_{12}v_1)^2
  (n_{12}v_2) v_2^4+455 (n_{12}v_1)
  (n_{12}v_2)^6\nonumber\\&\qquad-175 (n_{12}v_1) (n_{12}v_2)^4
  v_1^2-900 (n_{12}v_1) (n_{12}v_2)^4 (v_1v_2)-1400 (n_{12}v_1)
  (n_{12}v_2)^4 v_2^2\nonumber\\&\qquad+285 (n_{12}v_1) (n_{12}v_2)^2
  v_1^4+720 (n_{12}v_1) (n_{12}v_2)^2 v_1^2 (v_1v_2)+600 (n_{12}v_1)
  (n_{12}v_2)^2 v_1^2 v_2^2\nonumber\\&\qquad+600 (n_{12}v_1)
  (n_{12}v_2)^2 (v_1v_2)^2+1680 (n_{12}v_1) (n_{12}v_2)^2 (v_1v_2)
  v_2^2+1785 (n_{12}v_1) (n_{12}v_2)^2 v_2^4\nonumber\\&\qquad-1245
  (n_{12}v_1) v_1^6+180 (n_{12}v_1) v_1^4 (v_1v_2)-1980 (n_{12}v_1)
  v_1^4 v_2^2\nonumber\\&\qquad+360 (n_{12}v_1) v_1^2 (v_1v_2)^2-2475
  (n_{12}v_1) v_1^2 v_2^4+320 (n_{12}v_1)
  (v_1v_2)^3\nonumber\\&\qquad+300 (n_{12}v_1) (v_1v_2)^2 v_2^2+60
  (n_{12}v_1) (v_1v_2) v_2^4-2870 (n_{12}v_1)
  v_2^6\nonumber\\&\qquad+640 (n_{12}v_2)^7-276 (n_{12}v_2)^5
  v_1^2-266 (n_{12}v_2)^5 (v_1v_2)\nonumber\\&\qquad-2968
  (n_{12}v_2)^5 v_2^2+320 (n_{12}v_2)^3 v_1^4+100 (n_{12}v_2)^3 v_1^2
  (v_1v_2)\nonumber\\&\qquad+1300 (n_{12}v_2)^3 v_1^2 v_2^2+120
  (n_{12}v_2)^3 (v_1v_2)^2+560 (n_{12}v_2)^3 (v_1v_2)
  v_2^2\nonumber\\&\qquad+6000 (n_{12}v_2)^3 v_2^4-820 (n_{12}v_2)
  v_1^6+30 (n_{12}v_2) v_1^4 (v_1v_2)-1980 (n_{12}v_2) v_1^4
  v_2^2\nonumber\\&\qquad+240 (n_{12}v_2) v_1^2 (v_1v_2)^2-3960
  (n_{12}v_2) v_1^2 v_2^4+400 (n_{12}v_2)
  (v_1v_2)^3\nonumber\\&\qquad+720 (n_{12}v_2) (v_1v_2)^2 v_2^2+630
  (n_{12}v_2) (v_1v_2) v_2^4-9960 (n_{12}v_2)
  v_2^6\Bigr)v_{2}^i\Bigr]\,.
\end{align}
\end{subequations}

\bibliography{ListeRef_two-body-relat}

\end{document}